\documentclass[%
reprint,
superscriptaddress,
amsmath,amssymb,
aps,
pra,
]{revtex4-2}

\usepackage{mathrsfs}
\usepackage{graphicx}% Include figure files
\usepackage{dcolumn}% Align table columns on decimal point
\usepackage{bm}% bold math
\usepackage{amsmath, amssymb}
\usepackage{makecell}
\usepackage{multirow}
\usepackage{float}
\usepackage{color}
\usepackage{stmaryrd}
\usepackage{array}
\usepackage{setspace}
\usepackage{mathtools}
\def\opename{\operatorname}

\makeatletter

\newcommand{\Rmnum}[1]{\expandafter\@slowromancap\romannumeral #1@}
\makeatother                                                                   

\begin{document}                                                                
\preprint{APS/123-QED}
	
\title{Fourier modal method for perturbation analysis of bound states in the continuum} 
	
\author{Nan Zhang}
\email{nan25@vt.edu}
\affiliation{Bradley Department of Electrical and Computer Engineering, Virginia Tech, Blacksburg, Virginia 24061, USA}
\affiliation{Department of Mathematics, City University of Hong Kong, Kowloon, Hong Kong, China}
\author{Ya Yan Lu}%
\affiliation{Department of Mathematics, City University of Hong Kong, Kowloon, Hong Kong, China}
	
\date{\today}
% It is always \today, today, but any date may be explicitly specified
	
\begin{abstract}
Bound states in the continuum (BICs) supported by periodic photonic structures
are surrounded by high-$Q$ resonances 
and serve as centers of polarization vortices in momentum space, 
making them attractive for a wide range of applications. 
Recently, a perturbation theory has been formulated for resonant states near BICs, 
yielding a hierarchy of equations whose solutions directly approximate the resonant states 
and determine the asymptotic behavior of their $Q$ factors and polarizations. 
Efficiently solving these equations is therefore essential. 
Since many studies of BICs concern layered structures, 
we develop an approach based on the Fourier modal method.
Numerical examples confirm the effectiveness of the proposed method and 
demonstrate its ability to capture higher-order asymptotic scaling.
Our work provides a practical tool for the theoretical analysis of BICs 
and may also facilitate the design of BIC-based photonic devices.	
\end{abstract}

\maketitle
	
%\tableofcontents

%---------------------
\section{Introduction}
%---------------------

Photonic bound states in the continuum (BICs) are perfectly confined modes with frequencies embedded in the radiation continuum.
~\cite{Neumann1929PZ,Bonnet1994MMAS,Marinica08PRL,Bulgakov08PRB,Plotnik11PRL,Hsu13Nature,Hsu16NRM,Kivshar2023PU}.
They possess infinite radiative quality ($Q$) factors~\cite{Hsu13Nature}
and can be realized in a variety of optical systems,
including periodic structures~\cite{Hsu13Nature,Zhen14PRL}, waveguides~\cite{Webster07PTL,Zou15LPR,Zhang24OE}, and anisotropic media~\cite{Gomis17NP}.
In periodic structures such as photonic crystal (PhC) slabs,
a BIC is surrounded by a family of high-$Q$ resonant states
and can serve as the center of a far-field polarization vortex in momentum space~\cite{Zhen14PRL,Lee12PRL,Bulgakov17PRATopo}.
Owing to these unique properties, BICs have been widely explored 
for applications in lasing~\cite{Kodigala17Nature,Rong23NM}, 
quantum optics~\cite{Maria22Science,Rivera23PNAS}, 
structured-light generation~\cite{JZi20NP}, 
and chiral optics~\cite{Yuri20PRL,Alu21PRL,Pura24,Yuri22Science}.

Several theoretical frameworks have been proposed to elucidate fundamental aspects of BICs, 
including their physical origins and robustness against structural variations~\cite{Neumann1929PZ,Bonnet1994MMAS,Zhen14PRL,Friedrich85PRA,Yu26SIAMAM,Susumu14PRL,Kivshar19PRB,Yuan2017OL,Yuan2021PRAcond,Zhang23OE,Bezus24PRA}. 
Recently, a perturbation theory has been developed to analyze resonant states near a BIC in a periodic structure~\cite{Yuan20PRAPert,Zhang25PRL}. 
Expanding the resonant fields and frequencies in powers of a small wavevector perturbation parameter $\delta$
yields a sequence of perturbation equations. 
Their solutions provide reliable approximations to nearby resonant states, 
while the leading far-field radiation patterns determine the asymptotic behavior of the $Q$ factors and polarizations~\cite{Zhang25PRL}.

Efficiently solving the perturbation equations 
is therefore crucial for characterizing resonant states near a BIC. 
In some cases, higher-order solutions are required not only 
to improve the approximation of the resonant states, 
but also to capture the correct asymptotic behavior 
when the lower-order radiation patterns vanish. 
Super-BICs~\cite{Zhang25PRL,Yuan17PRA,Yuan2018PRA,Zhen19Nature,Bulgakov23PRB,Lee2023LPR,Shubin2023PRB,Le2024PRL,Zhang2025Super} provide an important example: 
the leading radiation pattern appears at order ${\mu}>1$, 
and the nearby resonances satisfy the asymptotic scaling $Q\sim Q_{2{\mu}}/\delta^{2{\mu}}$, 
in contrast to the generic case with ${\mu}=1$~\cite{Zhang25PRL,Yuan17PRA,Yuan2018PRA}. 
The resulting stronger enhancement of the $Q$ factor 
makes super-BICs particularly promising for applications 
such as ultralow-threshold lasers~\cite{Yuri21NC,Cui25NP}. 
These considerations motivate the development of an efficient recursive method 
for solving the perturbation equations to arbitrary order.

Given the widespread use of multilayer structures in theoretical 
and experimental studies of BICs~\cite{Hsu13Nature,Hsu16NRM,Kivshar2023PU,Zhen14PRL}, 
we develop an approach based on the Fourier modal method (FMM)~
\cite{Moharam1981,Li1996,Lalanne1996,Granet1996,Li1996josaa2,Whittaker1999,
Salakhova21PRB,Hao26PRB,Thomas26PRB,Nakagawa2002,Bai2007,Paul2010,Thomas2018OE} 
for computing perturbation solutions. 
A multilayer structure is periodic in the transverse plane and piecewise uniform along the longitudinal direction. 
In each layer, we treat the perturbation equations as source-driven problems,
reducing them to inhomogeneous ODEs for the Fourier coefficients.
We further represent the Fourier coefficients in terms of modal exponential vectors
and coefficient matrices, which admit a polynomial representation.
The recursive perturbation equations are thereby transformed into
a recursion among polynomial matrices, enabling efficient and stable computation
to arbitrary order.
Although methods for solving individual inhomogeneous ODE systems are well established~\cite{Nakagawa2002,Bai2007,Paul2010}, 
the present formulation is particularly suited to the recursive structure of the perturbation calculation. 
The solutions in all layers are coupled through a scattering-matrix formulation, 
while the frequency correction is obtained from either an analytical solvability condition or a discrete Fredholm condition. 
Numerical examples demonstrate that the method efficiently computes perturbation solutions and reproduces the higher-order behavior associated with super-BICs.

The remainder of this paper is organized as follows. 
Section~\Rmnum{2} derives the perturbation equations for resonant states near a BIC. 
Sections~\Rmnum{3} and \Rmnum{4} develop the new approach based on the FMM. 
Section~\Rmnum{5} presents numerical examples that validate the proposed method. 
Finally, Sec.~\Rmnum{6} presents the conclusions.

%------------------------------- 
\section{Perturbation equations} 
%------------------------------- 

In this section,
we derive the perturbation equations for resonant states
near a nondegenerate BIC.
The derivation extends the analysis in Ref.~\cite{Zhang25PRL}.
Specifically, we consider a nonmagnetic and lossless three-dimensional (3D) structure
that is biperiodic in the $xy$ plane with lattice vectors ${\bm l}_1$ and ${\bm l}_2$
and is sandwiched between two homogeneous isotropic media.
Throughout, we use Heaviside--Lorentz units with $c=1$ and assume an
$\exp(-i\omega t)$ time dependence.

Let ${\bm \Psi}_*=({\bm E}_*,{\bm H}_*)$, $\omega_*$, and  
${\bm\kappa}_*=(\alpha_*,\beta_*)$ denote the field, frequency, and  
in-plane Bloch wavevector of the BIC, respectively. 
Let ${\bm \Psi}=({\bm E},{\bm H})$, $\omega$, and ${\bm\kappa}$ denote  
the corresponding quantities of a nearby resonant state, with 
${\bm \kappa}  =  {\bm \kappa}_*  +  {\delta}{\bm \theta}/L$ 
and $0<\delta\ll 1$, 
where $L$ is the unit of length, ${\bm \theta}=(\cos\theta,\sin\theta)$ and  
$-\pi\leq\theta<\pi$.  
We expand $\omega$ and ${\bm \Psi}$ in powers of $\delta$ as  
\begin{align} 
	&\omega(\delta,\theta)  
	=  
	\omega_*  
	+  
	\sum_{j=1}^{\infty}  
	\delta^j\omega_j(\theta),  
	\label{eq:omegaExpansion} \\  
	&{\bm \Psi}({\bm r};\delta,\theta)  
	=  
	e^{i\delta{\bm\theta}\cdot{\bm\rho}/L}  
	\sum_{j=0}^{\infty}  
	\delta^j{\bm \Psi}_j({\bm r};\theta), 
	\label{eq:Fexpansion} 
\end{align}  
where ${\bm\rho}=(x,y)$ and ${\bm \Psi}_0={\bm \Psi}_*$.  
The fields ${\bm \Psi}_j$ have the same quasi-periodicity  
as the BIC.  

Substituting Eqs.~\eqref{eq:omegaExpansion} and \eqref{eq:Fexpansion} 
into Maxwell's equations ${\cal A}{\bm \Psi}  = \omega{\cal B}{\bm \Psi}$,
where  
\begin{equation}  
	{\cal A}  
	=  
	\begin{bmatrix}  
		 & i\nabla\times\\  
		-i\nabla\times & 
	\end{bmatrix},  
	\quad  
	{\cal B}  
	=  
	\begin{bmatrix}  
		{\bm\varepsilon}({\bm r}) & \\  
		 & {\bm I}  
	\end{bmatrix}, 
	\label{eq:operatorsAandB} 
\end{equation}  
and collecting terms of the same order in $\delta$, 
we obtain  
\begin{equation}  
	({\cal A}-\omega_*{\cal B}){\bm \Psi}_j  
	=  
	{\bm f}_j,\quad j\geq 0,
	\label{eq:MaxwellPerturbationEquation}
\end{equation}  
where ${\bm f}_0={\bm 0}$, 
so that $j=0$ corresponds to the BIC, and
\begin{equation}  
	{\bm f}_j  
	=  
	-{\cal A}_1(\theta){\bm \Psi}_{j-1}  
	+  
	\sum_{\ell=1}^{j}  
	\omega_\ell(\theta){\cal B}{\bm \Psi}_{j-\ell},\quad j\geq1, 
	\label{eq:MaxwellPerturbationSource}
\end{equation}  
with
\begin{equation} 
	{\cal A}_1(\theta) 
	= 
	\frac{1}{L} 
	\begin{bmatrix} 
		{\bm 0} & -{\bm\theta}\times\\ 
		{\bm\theta}\times & {\bm 0} 
	\end{bmatrix}. 
	\label{eq:MaxwellA1} 
\end{equation} 
For vector fields  
${\bm X}$ and ${\bm Y}$ and a linear operator ${\cal L}$, we define  
\begin{equation}  
	\langle{\bm X}|{\cal L}|{\bm Y}\rangle  
	=  
	\frac{1}{4L^3}  
	\int_\Omega  
	{\bm X}^{\dagger}({\bm r})  
	{\cal L}{\bm Y}({\bm r})  
	\,{\rm d}{\bm r}.  
	\label{eq:fieldInnerProduct}  
\end{equation}  
Here ``$\dagger$'' denotes the conjugate transpose and 
$\Omega$ is a unit cell. 
We normalize the BIC such that  
$\langle{\bm \Psi}_*|{\cal B}|{\bm \Psi}_*\rangle=1$ and impose the gauge condition
$\langle{\bm \Psi}_*|{\cal B}|{\bm \Psi}_j\rangle=0$ for $j\geq1$.  
The solvability condition for Eq.~\eqref{eq:MaxwellPerturbationEquation}  
then gives  
\begin{equation}  
	\omega_j(\theta)  
	=  
	\langle{\bm \Psi}_*|{\cal A}_1(\theta)|{\bm \Psi}_{j-1}\rangle.  
	\label{eq:omegaCorrection}  
\end{equation}

We next examine the far-field radiation patterns of the perturbation 
fields and derive the asymptotic behavior of the $Q$ factor and 
polarizations. We assume that only the zeroth diffraction order 
propagates in each of the upper and lower homogeneous media. Let 
$\varepsilon^+$ and $\varepsilon^-$ denote the dielectric constants 
of the upper and lower media, respectively. The radiation patterns of 
the resonant field have the form 
\begin{equation} 
{\bm \Psi} 
	\sim 
    {\bm \psi}^\pm
	e^{i{\bm k}^{\pm}\cdot{\bm r}}, 
	\quad 
	z\to\pm\infty, 
\end{equation} 
where ${\bm k}^{\pm}=({\bm\kappa},\pm\gamma^\pm)$ and 
$\gamma^\pm=\sqrt{\omega^2\varepsilon^\pm-|{\bm\kappa}|^2}$. 
The square roots are chosen according to the outgoing radiation 
condition.  
The longitudinal wavenumbers $\gamma^\pm$ also depend on $\delta$, 
and the radiation patterns of the perturbation fields generally 
contain polynomial factors in $z$. More precisely, 
\begin{equation} 
	{\bm \Psi}_j({\bm r}) 
	\sim 
    {\bm \psi}_j^\pm(z)	
	e^{i{\bm k}_*^\pm\cdot{\bm r}}, 
	\quad 
	z\to\pm\infty, 
\end{equation} 
where ${\bm k}_*^\pm=({\bm\kappa}_*,\pm\gamma_*^\pm)$, 
$\gamma_*^\pm=\sqrt{\omega_*^2\varepsilon^\pm 
	-|{\bm\kappa}_*|^2}>0$, and 
${\bm \psi}_j^\pm(z)$ are vector-valued polynomials in 
$z$, respectively.

For a given angle $\theta$, 
let $\mu$ denote the first nonvanishing order of the far-field radiation, i.e.,
${\bm \psi}_m^\pm(z)=0$ for $m<\mu$ and ${\bm \psi}_\mu^\pm(z)\neq 0$.
We can show that
${\bm \psi}_\mu^\pm(z)$ reduce to the constant vectors
$({\bm e}_\mu^\pm,{\bm h}_\mu^\pm)$.
Correspondingly, 
\begin{equation} 
	{\bm \psi}^\pm=\begin{bmatrix} 
		{\bm e}^{\pm}\\ 
		{\bm h}^{\pm} 
	\end{bmatrix}
	= 
	\delta^\mu	\begin{bmatrix} 
		{\bm e}_\mu^{\pm}\\ 
		{\bm h}_\mu^{\pm}
	\end{bmatrix}
	+ 
	O(\delta^{\mu+1}).
\end{equation} 
From the terms of order 
$\delta^\mu$ in 
${\bm k}^\pm\cdot {\bm e}^\pm={\bm k}^\pm\cdot {\bm h}^\pm=0$, we obtain 
${\bm k}_*^\pm\cdot{\bm e}_\mu^\pm={\bm k}_*^\pm\cdot{\bm h}_\mu^\pm=0$. 
{\em Therefore, the leading polarization lies in the plane 
perpendicular to ${\bm k}_*^\pm$. }

Applying the complex Poynting theorem to Maxwell's equations,
as detailed in Appendix~\ref{appe:derEqQasym}, we find that the $Q$
factor, defined by $Q=-0.5{\opename{Re}(\omega)}/{\opename{Im}(\omega)}$,
satisfies $Q(\delta)\sim{Q_{2\mu}}/{\delta^{2\mu}}$
where
\begin{equation}
	Q_{2\mu}
	=
	\frac{\omega_*L}
	{{\cal P}_\mu^++{\cal P}_\mu^-},
	\label{eq:Q2mu}
\end{equation}
and
\begin{equation}
	{\cal P}_\mu^\pm
	=
	\pm\frac{A}{2}
	\opename{Re}\!\left(
	{\bm e}_\mu^\pm
	\times
	\overline{{\bm h}_\mu^\pm}
	\right)\cdot\hat{z}.
	\label{eq:Pmu}
\end{equation}
Here 
$A=\left|\det{\bm L}_{\rho}\right|/L^2$ 
is the dimensionless area of the lattice cell with
${\bm L}_{\rho}=[{\bm l}_1,{\bm l}_2]$ and 
${\cal P}_\mu^\pm$ are the 
powers radiated by the $\mu$-th order field 
into the upper and lower media, respectively. {\em The physical meaning 
of the above result is transparent: the asymptotic order of the 
$Q$ factor is determined by the order of the first nonzero radiative 
field, while, under the chosen normalization of the BIC energy, the 
corresponding asymptotic coefficient is completely determined by the outgoing powers.}

For a generic BIC, $\mu=1$ for all $\theta$. When $\mu>1$ for some 
$\theta$, the BIC is a super-BIC, and the vanishing of the radiation 
patterns up to order $\mu-1$ leads to the corresponding higher-order 
scaling. 

We finally note a useful structural property of the perturbation 
solutions. Since the perturbation expansion is generally analytic in 
the wavevector detuning, for each $j\geq1$, $\omega_j(\theta)$ and 
${\bm \Psi}_j({\bm r};\theta)$ are homogeneous polynomials of degree $j$ 
in the components of ${\bm\theta}$. Specifically, 
\begin{align} 
	&\omega_j(\theta) 
	= 
	\sum_{m=0}^{j} 
	\omega_{j,m} 
	(\cos\theta)^m 
	(\sin\theta)^{j-m}, 
	\\ 
	&{\bm \Psi}_j({\bm r};\theta) 
	= 
	\sum_{m=0}^{j} 
	{\bm \Psi}_{j,m}({\bm r}) 
	(\cos\theta)^m 
	(\sin\theta)^{j-m}, 
\end{align} 
where $\omega_{j,m}$ and ${\bm \Psi}_{j,m}({\bm r})$ are independent of 
$\theta$. Consequently, at each order $j$, it is sufficient to solve 
the perturbation equations along $j+1$ suitably chosen directions to 
determine $\omega_j(\theta)$ and ${\bm \Psi}_j({\bm r};\theta)$ for 
arbitrary $\theta$. 

The leading far-field radiation amplitudes inherit the same polynomial structure. 
In particular, if the first $\mu-1$ perturbation fields are nonradiating for all $\theta$, 
then
\begin{equation} 
	\begin{bmatrix} 
		{\bm e}_\mu^{\pm}(\theta)\\ 
		{\bm h}_\mu^{\pm}(\theta) 
	\end{bmatrix}
	= 
	\sum_{m=0}^{\mu} 
	\begin{bmatrix} 
		{\bm e}_{\mu,m}^{\pm}\\ 
		{\bm h}_{\mu,m}^{\pm}
	\end{bmatrix}
	(\cos\theta)^m 
	(\sin\theta)^{\mu-m}.
	\label{eq:ehmuthetahomo} 
\end{equation} 
Thus, the leading radiation amplitudes are  
determined by their values along $\mu+1$ suitably chosen directions. 
Additional spatial symmetries can further reduce the number of 
directions required. 

The theory developed above applies to arbitrary 3D periodic 
structures, and the perturbation equations can, in principle, be 
solved using standard numerical methods~\cite{monk2003finite,Zhang26JCP}. For layered structures, 
however, we next present an FMM-based algorithm for computing 
the frequency corrections 
$\omega_j$ and the 
perturbation solutions ${\bm \Psi}_j$.

%--------------------------------------- 
\section{FMM for source-driven problems} 
%--------------------------------------- 

The perturbation equations derived in the preceding section can be 
regarded as source-driven Maxwell problems.  In this section, we 
review the FMM formulation for such problems.  The recursive method 
for solving the perturbation equations is presented in the subsequent 
sections. 

We assume that the structure is layered and piecewise uniform in 
$z$.  The first and last layers, labeled $1$ and $P$, are 
semi-infinite homogeneous regions.  A finite layer indexed by 
$1<p<P$ occupies the region $z_p<z<z_p+d_p$.  Within layer $p$, 
the relative permittivity tensor is written as 
${\bm\varepsilon}_p({\bm\rho})$.   

We consider Maxwell's equations with an electric current density 
${\bm J}$ and a magnetic current density ${\bm M}$. 
Let $f$ denote any Cartesian 
component of ${\bm E}$, ${\bm H}$, ${\bm J}$, or ${\bm M}$.  It can be 
expanded as 
\begin{equation} 
	f({\bm r}) 
	= 
	\sum_{{\bm G}} 
	f_{{\bm G}}(z) 
	e^{i({\bm\kappa}+{\bm G})\cdot{\bm\rho}}, 
\end{equation} 
where ${\bm G}\in{\bm L}_G\mathbb Z^2$ is a reciprocal-lattice vector,
${\bm L}_G=2\pi{\bm L}_\rho^{-{\sf T}}$ is the reciprocal-lattice matrix,
and $f_{{\bm G}}(z)$ is the corresponding Fourier coefficient.
After truncating to
$N$ Fourier harmonics, the coefficients of the electric and magnetic
fields and currents are collected into the $3N$-dimensional vectors
${\bf E}$, ${\bf H}$, ${\bf J}$, and ${\bf M}$, respectively.
We denote by ${\bm \Sigma}(z)$ the $4N$-dimensional column vector formed by the Fourier
coefficients of the tangential electromagnetic field components.
In each layer, Maxwell's equations are reduced to a first-order system 
of inhomogeneous ODEs: 
\begin{equation}
	\frac{{\rm d}{\bm\Sigma}}{{\rm d}z}
	-
	i{\bm \Pi}_{p}{\bm\Sigma}
	=
	{\bf q}_{p}(z),
	\label{eq:systemode}
\end{equation}
where the source vector ${\bf q}_{p}$ is expressed as
\begin{equation}
	{\bf q}_p(z)
	=
	{\bf Q}_{{\rm J},p}{\bf J}(z)
	+
	{\bf Q}_{{\rm M},p}{\bf M}(z).
	\label{eq:qJMrelation}
\end{equation}
The matrices appearing above are given in Appendix~\ref{appe:explicitofmatrices}.

We solve Eq.~\eqref{eq:systemode} using an eigenmode expansion. 
We first introduce a local coordinate ${\xi}$ defined by 
$\xi=z-z_{p}$ for $1<{p}\leq {P}$ and ${\xi}=z-z_2$ for ${p}=1$. 
All quantities in layer ${p}$ are henceforth expressed as functions of ${\xi}$. 
We expand the field vector in the eigenmodes of layer ${p}$ as 
${\bm \Sigma}({\xi})={\bf W}_{p}{\bf c}_{p}({\xi})$, 
where ${\bf W}_{p}$ satisfies the eigenvalue problem 
${\bm \Pi}_{p}{\bf W}_{p}={\bf W}_{p}{\bm \Gamma}_{p}$. 
The eigenvalue matrix ${\bm \Gamma}_p$ has the block-diagonal form
${\bm \Gamma}_p=\opename{diag}\left(
	{\bm \Gamma}_p^+,
	{\bm \Gamma}_p^-\right)$,
where the two diagonal blocks are
\begin{align}
	{\bm \Gamma}_p^\pm
	&=
	\opename{diag}
	\left(
	\gamma_{p,1}^\pm,\ldots,
	\gamma_{p,2N}^\pm
	\right).
\end{align}
The superscripts ``$\pm$'' label modes
directed toward increasing and decreasing $z$, respectively.

Substituting the modal expansion into Eq.~\eqref{eq:systemode} gives 
\begin{equation} 
	\frac{{\rm d}{\bf c}_p}{{\rm d}\xi} 
	- 
	i{\bm \Gamma}_p{\bf c}_p 
	= 
	{\bf g}_p, 
	\quad 
	{\bf g}_p 
	= 
	{\bf W}_{p}^{-1}{\bf q}_p. 
	\label{eq:c_equation} 
\end{equation} 
In a finite layer $1<p<P$, we decompose the modal coefficient 
vector as 
\begin{equation} 
	{\bf c}_p(\xi) 
	= 
	{\bf U}_p(\xi) 
	\left[ 
	\begin{array}{c} 
		{\bf a}_p\\ 
		{\bf b}_p 
	\end{array} 
	\right]
	+ 
	{\bf r}_p(\xi), 
	\label{eq:modal-general-solution} 
\end{equation} 
where ${\bf r}_p$ is a particular 
solution of Eq.~\eqref{eq:c_equation}
and
\begin{equation}
	{\bf U}_p(\xi)
	=
	\opename{diag}
	\left[
	{\bf U}_p^+(\xi),
	{\bf U}_p^-(\xi-d_p)
	\right],
	\label{eq:general-propagation-matrix}
\end{equation}
is the propagation matrix, where
\begin{align}
	{\bf U}_p^\pm(\xi)
	&
	=
	\opename{diag}
	\left(
	e^{i\gamma_{p,1}^\pm\xi},\ldots,
	e^{i\gamma_{p,2N}^\pm\xi}
	\right).
\end{align}
The vectors 
${\bf a}_p$ and ${\bf b}_p$ contain the amplitudes of modes directed 
toward increasing and decreasing $\xi$. 
The same representation is used for the exterior layers by setting $d_1=d_{P}=0$.

%We introduce a source injection vector ${\bf w}_{p}$ in each layer. 
%For a finite layer ${p}$, we define
%\begin{equation} 
%	{\bf w}_p 
%	= 
%	\left[ 
%	\begin{array}{c} 
%		{\bf r}_p^+(d_p) 
%		- 
%		{\bm \Lambda}_p^+{\bf r}_p^+(0) 
%		\\ 
%		{\bf r}_p^-(0) 
%		- 
%		{\bm \Lambda}_p^-{\bf r}_p^-(d_p) 
%	\end{array} 
%	\right]. 
%	\label{eq:general-source-injection} 
%\end{equation} 
%where ${\bm \Lambda}_p^{\pm} = {\bf U}_p^{\pm}(\pm d_p)$.
%For the exterior layers, we define
%\begin{equation} 
%	{\bf w}_1 
%	= 
%	{\bf W}_1 
%	\left[ 
%	\begin{array}{c} 
%		{\bf r}_1^+(0)\\ 
%		{\bf r}_1^-(0) 
%	\end{array} 
%	\right], 
%	\quad 
%	{\bf w}_P 
%	= 
%	-{\bf W}_P 
%	\left[ 
%	\begin{array}{c} 
%		{\bf r}_P^+(0)\\ 
%		{\bf r}_P^-(0) 
%	\end{array} 
%	\right]. 
%\end{equation} 
Finally, 
we impose the interface continuity conditions and use the scattering matrix formulas~\cite{Li1996josaa2,Whittaker1999} to determine the amplitude vectors. 
The scattering relation in its standard form is given by
\begin{equation}\label{eq:scatteringformula}
	\left[
	\begin{array}{c}
		{\bf a}_{P}\\
		{\bf b}_{1}
	\end{array}
	\right]
	=
	{\bf S}
	\left[
	\begin{array}{c}
		{\bf a}_{1}\\
		{\bf b}_{P}
	\end{array}
	\right]
	+
	{\bf s},
\end{equation}
where ${\bf S}$ is the scattering matrix and ${\bf s}$ is the source contribution arising from the particular solutions ${\bf r}_{p}$ in all layers.
The layer-by-layer recursive procedure for constructing ${\bf S}$ and ${\bf s}$ can be found in Ref.~\cite{Li1996josaa2,Whittaker1999}.
Equivalently, the inverse form of Eq.~\eqref{eq:scatteringformula} is
\begin{equation}
	\left[
	\begin{array}{c}
		{\bf a}_{1}\\
		{\bf b}_{P}
	\end{array}
	\right]
	=
	{\bf S}^{-1}
	\left[
	\begin{array}{c}
		{\bf a}_{P}\\
		{\bf b}_{1}
	\end{array}
	\right]
	+
	{\bf t}.
\end{equation}
Both ${\bf S}^{-1}$ and ${\bf t}$ can be obtained directly through a layer-by-layer recursive procedure analogous to that used for the standard form.

For the source-free problem, ${\bf s}={\bf t}={\bf 0}$. 
A resonant state satisfies the outgoing radiation condition, 
which requires the incoming modal amplitudes ${\bf a}_1$ and ${\bf b}_{P}$ to vanish. 
Thus, a resonant state exists when ${\bf S}^{-1}$ is singular, 
corresponding to a pole of the meromorphic continuation of ${\bf S}$.
A BIC is such a state at a real frequency and a real Bloch wavevector, 
for which all propagating components of ${\bf a}_{P}$ and ${\bf b}_1$ vanish, 
leaving only evanescent modes in the exterior layers.

%-----------------------------------------------------------
\section{Recursive algorithm for the perturbation equations}
%-----------------------------------------------------------

In this section, we develop a stable and efficient recursive algorithm 
for solving the hierarchy of perturbation equations. 
At each order, we regard the perturbation equation 
as a source-driven problem with 
electric and magnetic currents ${\bm J}_j$ and ${\bm M}_j$ 
that are formed from the lower-order solutions.
Building on the formulation developed in the preceding section, 
the main task is to recursively construct and solve the inhomogeneous ODE systems~\eqref{eq:c_equation}. 
To facilitate the recursion, 
we decompose the Fourier coefficient vectors of the fields and 
sources into products of polynomial matrices and a common modal exponential vector, 
so that the perturbation equations can be reduced to recursive relations among these polynomial matrices.
In the remainder of this section, we present the details of the method.

For notational simplicity, the layer index $p$ is suppressed below.
Let ${\bm \varphi}(\xi)$ denote the vector of modal exponential
functions, i.e., the diagonal elements of ${\bf U}(\xi)$. At
perturbation order $j$, we write the Fourier coefficient vectors of fields and currents as
\begin{equation}
	{\bf E}_j(\xi)
	=
	{\mathbb E}_j(\xi)
	{\bm \varphi}(\xi),
	\label{eq:Ejrepresentation}
\end{equation}
with analogous representations for ${\bf H}_j$, ${\bf J}_j$, and
${\bf M}_j$. The modal source vector and the particular solution are
represented similarly as
\begin{equation}
	{\bf g}_j(\xi)
	=
	{\mathbb G}_j(\xi)
	{\bm \varphi}(\xi),
	\quad
	{\bf r}_j(\xi)
	=
	{\mathbb R}_j(\xi)
	{\bm \varphi}(\xi).
	\label{eq:gjrepresentation}
\end{equation}
We also write the modal coefficient vector as
${\bf c}_j(\xi)={\mathbb C}_j(\xi){\bm \varphi}(\xi)$.
These matrices denoted by blackboard bold symbols are polynomial in $\xi$.

For the BIC at $j=0$, 
since there are no sources at this order, 
${\mathbb J}_{0}={\mathbb M}_{0}={\mathbb G}_{0}={\mathbb R}_{0}={\bf 0}$. 
The modal coefficient matrix ${\mathbb C}_{0}$ is determined from the homogeneous FMM solution, 
from which the field coefficient matrices ${\mathbb E}_{0}$ and ${\mathbb H}_{0}$ are reconstructed. 
These matrices are all constant, establishing the polynomial representation at zeroth order.

At order $j\geq1$, suppose that the frequency corrections and
perturbation fields have been determined up to order $j-1$. The
correction $\omega_j$ is first determined from the solvability
condition~(\ref{eq:omegaCorrection}). The Fourier representations and
polynomial structure of the lower-order perturbation fields allow
$\omega_j$ to be evaluated analytically.

Let ${\bf A}_1$ and ${\bf B}$ denote the Fourier matrix
representations of ${\cal A}_1$ and ${\cal B}$, respectively. The
electric and magnetic current coefficient matrices satisfy
\begin{equation}
	i
	\left[
	\begin{array}{c}
		{\mathbb J}_j\\
		{\mathbb M}_j
	\end{array}
	\right]
	=
	-
	{\bf A}_1
	\left[
	\begin{array}{c}
		{\mathbb E}_{j-1}\\
		{\mathbb H}_{j-1}
	\end{array}
	\right]
	+
	\sum_{\ell=1}^{j}
	\omega_\ell
	{\bf B}
	\left[
	\begin{array}{c}
		{\mathbb E}_{j-\ell}\\
		{\mathbb H}_{j-\ell}
	\end{array}
	\right].
	\label{eq:JMmatrixrecursive}
\end{equation}
Following Eqs.~\eqref{eq:qJMrelation} and~\eqref{eq:c_equation}, the
modal source matrix is
\begin{equation}
	{\mathbb G}_j
	=
	{\bf W}^{-1}{\bf Q}_{\rm J}{\mathbb J}_j
	+
	{\bf W}^{-1}{\bf Q}_{\rm M}{\mathbb M}_j.
	\label{eq:Gjrecursive}
\end{equation}
Since the field coefficient matrices are
polynomial matrices, ${\mathbb J}_j$, ${\mathbb M}_j$, and
${\mathbb G}_j$ are also polynomial matrices.

Substituting Eq.~\eqref{eq:gjrepresentation} into
Eq.~\eqref{eq:c_equation} gives
\begin{equation}
	\frac{{\rm d}{\mathbb R}_j}{{\rm d}\xi}
	+
	i
	\left(
	{\mathbb R}_j{\bm \Gamma}
	-
	{\bm \Gamma}{\mathbb R}_j
	\right)
	=
	{\mathbb G}_j.
	\label{eq:Rmatrixode}
\end{equation}
A stable algorithm for solving Eq.~\eqref{eq:Rmatrixode} 
is developed in the next subsection, 
which ensures that ${\mathbb R}_j$ remains a polynomial matrix. 
This algorithm is in fact the key step in preserving the polynomial 
structure of all coefficient matrices throughout the recursive calculation.

In each layer, 
once ${\mathbb R}_{j,p}$ has been obtained, 
the particular solutions ${\bf r}_{j,p}$ enter the interface
continuity conditions as known inhomogeneous terms. 
These contributions are combined through the scattering-matrix procedure 
to obtain the global source vector ${\bf s}_j$.
Since the perturbation field
satisfies the outgoing radiation condition, the incoming modal
amplitudes vanish, and hence
\begin{equation}
	\left[
	\begin{array}{c}
		{\bf a}_{j,P}\\
		{\bf b}_{j,1}
	\end{array}
	\right]
	=
	{\bf s}_j.
	\label{eq:scatteringPerturbation}
\end{equation}
The modal amplitudes ${\bf a}_{j,p}$ and ${\bf b}_{j,p}$ in the
interior layers are then recovered through the layer-by-layer
back-substitution procedure.

Together with the particular solutions, these modal amplitudes
determine the perturbation field at order $j$ up to an arbitrary
multiple of the BIC field. This remaining freedom is removed by
imposing the gauge condition introduced in Sec.~\Rmnum{2}. 
Suppressing the layer index again,
the complete modal coefficient matrix is then
\begin{equation}
	{\mathbb C}_{j}(\xi)
	=
	\operatorname{diag}
	\left(
	\left[
	\begin{array}{c}
		{\bf a}_{j}\\
		{\bf b}_{j}
	\end{array}
	\right]
	\right)
	+
	{\mathbb R}_{j}(\xi).
	\label{eq:Cjcomplete}
\end{equation}
Using the field reconstruction relations of the FMM, the field
coefficient matrices are obtained from
\begin{equation}
	{\mathbb E}_{j}
	=
	{\bf T}_{{\rm E}}{\mathbb C}_{j}
	+
	{\bf T}_{{\rm EJ}}{\mathbb J}_{j},
	\quad
	{\mathbb H}_{j}
	=
	{\bf T}_{{\rm H}}{\mathbb C}_{j}
	+
	{\bf T}_{{\rm HM}}{\mathbb M}_{j}.
	\label{eq:CJtoEH}
\end{equation}
The constant matrices ${\bf T}_{{\rm E}}$, ${\bf T}_{{\rm EJ}}$,
${\bf T}_{{\rm H}}$, and ${\bf T}_{{\rm HM}}$ are given in
Appendix~\ref{appe:explicitofmatrices}. 
Consequently,
${\mathbb C}_j$, ${\mathbb E}_{j}$ and ${\mathbb H}_{j}$ retain the polynomial
structure. 
The perturbation fields
${\bm \Psi}_j=({\bm E}_j,{\bm H}_j)$ are thereby obtained, completing
the solution at order $j$.
These field coefficient matrices are then used in Eq.~\eqref{eq:JMmatrixrecursive} 
to construct the sources at order $j+1$, 
allowing the recursion to proceed to the next order.

Although general methods are available for solving an individual
inhomogeneous ODE system~\eqref{eq:c_equation}, the present formulation
is particularly suited to the recursive perturbation hierarchy.
First, it converts the recursion among the perturbation equations
directly into a recursion among polynomial matrices, thereby facilitating
their construction and solution.
Second, the solution of either Eq.~\eqref{eq:c_equation} or
Eq.~\eqref{eq:Rmatrixode} requires special treatment when two modal
propagation constants $\gamma_m$ and $\gamma_n$ are close~\cite{Paul2010}.
Such cases require particular care in recursive calculations, since
numerical errors may propagate to higher perturbation orders.
As shown below, the proposed algorithm provides a systematic and
convenient way to handle these nearly degenerate cases within the
recursive procedure.

%-----------------------------------------------------------
\subsection{Stable calculation of Eq.~\eqref{eq:Rmatrixode}}
%-----------------------------------------------------------

Here we develop a stable procedure for solving
Eq.~\eqref{eq:Rmatrixode}. We formulate the procedure for a finite layer,
although the same algorithm also applies to the semi-infinite exterior layers.
For notational simplicity, the perturbation-order and layer indices are suppressed.
We introduce the normalized coordinate
$\zeta=\xi/d$
and the source matrix is represented as a matrix polynomial in $\zeta$,
\begin{equation}
	{\mathbb G}(\zeta)
	=
	\sum_{k=0}^{K}
	\zeta^k{\mathbb G}_k,
	\label{eq:Gpolynomial}
\end{equation}
where $K$ denotes the degree.

Since ${\bm \Gamma}$ is diagonal,
Eq.~\eqref{eq:Rmatrixode} can be solved elementwise. Let $\gamma_m$
and $\gamma_n$ denote the $m$-th and $n$-th diagonal entries of
${\bm \Gamma}$, respectively. 
For fixed $m$ and $n$, we denote
$r(\zeta)=[{\mathbb R}(\zeta)]_{mn}$,
$g_k=[{\mathbb G}_k]_{mn}$, and
$\Delta=d(\gamma_n-\gamma_m)$.
Equation~\eqref{eq:Rmatrixode} then reduces to
\begin{equation}
	\frac{{\rm d}r}{{\rm d}\zeta}
	+
	i\Delta r
	=
	d\sum_{k=0}^{K}\zeta^k g_k.
	\label{eq:Rentry}
\end{equation}

Let $\eta>0$ be a prescribed threshold. For $|\Delta|\geq\eta$, we
seek a polynomial particular solution
\begin{equation}
	r(\zeta)
	=
	\sum_{k=0}^{K}
	\zeta^k r_k.
\end{equation}
Substitution into Eq.~\eqref{eq:Rentry} and matching equal powers of
$\zeta$ gives
\begin{equation}
	r_k
	=
	\frac{
		dg_k-(k+1)r_{k+1}
	}{
		i\Delta
	}.
	\label{eq:downwardrecursion}
\end{equation}
Starting from $r_{K+1}=0$, the coefficients are evaluated downward
for $k=K,\ldots,0$.

For $|\Delta|<\eta$, direct use of
Eq.~\eqref{eq:downwardrecursion} may become numerically unstable.
We instead use the power series
\begin{equation}
	r(\zeta)
	=
	\sum_{k=0}^{\infty}
	\zeta^k r_k,
	\quad
	r_0=0.
\end{equation}
Matching equal powers of $\zeta$ gives
\begin{equation}
	r_{k+1}
	=
	\frac{
		dg_k-i\Delta r_k
	}{
		k+1
	},
	\label{eq:upwardrecursion}
\end{equation}
where $g_k=0$ for $k>K$. The coefficients are evaluated successively
from $k=0$.

Let $\epsilon_{\rm abs}>0$ and $\epsilon_{\rm rel}>0$ be prescribed
absolute and relative tolerances, respectively. For $q\geq K$, the
upward recursion is terminated at the first value of $q$ satisfying
\begin{equation}
	|r_{q+1}|
	<
	\epsilon_{\rm abs}
	+
	\epsilon_{\rm rel}
	\sum_{k=0}^{q}|r_k|.
\end{equation}
Thus, even in the nearly degenerate case with $|\Delta|\ll 1$,
the polynomial structure is retained to high numerical accuracy.
In the numerical examples given in Sec.~\Rmnum{5}, we take
$\eta=1$ and
$\epsilon_{\rm abs}=\epsilon_{\rm rel}=10^{-14}$,
which is sufficient to carry out the recursion reliably to the required perturbation order.
The case $\Delta=0$ is included automatically, 
and $r_k=0$ for all $k>K+1$.

%-----------------------------------------------------------
\subsection{Fredholm verification of $\omega_j$}
%-----------------------------------------------------------

The value of $\omega_j$ obtained from the analytical solvability
condition~\eqref{eq:omegaCorrection} can be independently verified
using a Fredholm solvability condition for the global inverse
scattering relation.
For this purpose, we separate the dependence of the current matrices
on $\omega_j$ as
\begin{equation}
	\left[
	\begin{array}{c}
		{\mathbb J}_j\\
		{\mathbb M}_j
	\end{array}
	\right]
	=
	\left[
	\begin{array}{c}
		{\mathbb J}_j^0\\
		{\mathbb M}_j^0
	\end{array}
	\right]
	+
	\omega_j
	\left[
	\begin{array}{c}
		{\mathbb J}_{\omega}\\
		{\mathbb M}_{\omega}
	\end{array}
	\right].
	\label{eq:JMsplit}
\end{equation}
According to Eqs.~\eqref{eq:MaxwellPerturbationSource} and~\eqref{eq:JMmatrixrecursive}, 
the matrices appearing in the above equation can be written explicitly.
The corresponding particular solutions can then be calculated.
Combining their contributions through the inverse scattering procedure 
gives the global source vectors 
${\bf t}_j^0$ and ${\bf t}_{\omega}$.
Since the incoming modal amplitudes vanish, the global inverse
scattering relation becomes
\begin{equation}
	{\bf S}^{-1}
	\left[
	\begin{array}{c}
		{\bf a}_{j,P}\\
		{\bf b}_{j,1}
	\end{array}
	\right]
	=
	-
	{\bf t}_j^0
	-
	\omega_j{\bf t}_{\omega}.
	\label{eq:inverseScatteringFredholm}
\end{equation}
The matrix ${\bf S}^{-1}$ is evaluated at the BIC frequency
and Bloch wavevector and is therefore singular.
Let ${\bf l}$ be a left null vector of ${\bf S}^{-1}$.
If ${\bf l}^{\dagger}{\bf t}_{\omega}\neq0$, the Fredholm
solvability condition gives
\begin{equation}
	\omega_j
	=
	-
	\frac{
		{\bf l}^{\dagger}{\bf t}_j^0
	}{
		{\bf l}^{\dagger}{\bf t}_{\omega}
	}.
	\label{eq:omegaFredholm}
\end{equation}
Agreement between Eq.~\eqref{eq:omegaFredholm} and the analytical
solvability condition~\eqref{eq:omegaCorrection} provides an
independent verification of both the frequency correction $\omega_j$
and the recursive FMM implementation.

%-----------------------------------------------------------
\section{Numerical Examples}
%---------------------------

In this section, we present three numerical examples 
to validate the proposed algorithm. 
The first example considers a diffraction grating 
and solves the perturbation equations up to eighth order 
for an at-$\Gamma$ BIC and an off-$\Gamma$ BIC, 
demonstrating the efficient computation of high-order perturbation solutions 
and providing increasingly accurate approximations of nearby resonant states. 
The second and third examples apply the algorithm to 
super-BICs with ${\mu}=3$ and ${\mu}=2$ in PhC slabs 
with square and triangular lattices, respectively. 
The computed high-order radiation patterns 
reproduce the correct asymptotic behavior 
of the polarization and the $Q$ factor 
when the lower-order radiation patterns vanish.

All numerical calculations are performed using a self-consistent truncated FMM formulation.
For a given number $N$ of retained harmonics, 
the BIC, nearby resonant states, and perturbation solutions are all computed with the same truncation.
Therefore, the very small errors, such as those of order $10^{-13}$ shown in Fig.~\ref{ex1},
measure the agreement between the perturbation solutions and the directly computed resonant states
within the same truncated model, 
rather than the absolute accuracy of the FMM discretization.

\subsection{BICs in diffraction gratings}
We consider a diffraction grating that is invariant in $x$
and periodic in $y$ with period $L$.
It is a two-dimensional (2D) structure consisting of
a finite grating layer embedded in air.
The grating layer has a height of $1.25L$,
and each period contains two dielectric regions,
each of width $0.5L$,
with refractive indices $1.45$ and $2.55$, respectively.
Although the algorithm developed in this paper is formulated
for 3D structures,
it can be applied to this 2D structure with only minor modifications.

The structure supports two TE BICs with $\alpha_*=0$. 
The first is an at-$\Gamma$ BIC with propagation constant $\beta_*=0$ 
and frequency $\omega_*L/2\pi=0.640$, 
while the second is an off-$\Gamma$ BIC with 
$\beta_*L/2\pi=0.181$ and $\omega_*L/2\pi=0.623$. 
For each BIC, we solve the associated perturbation equations up to eighth order. 
The order-$j$ perturbation approximation to the resonant frequency, 
denoted by $\omega_{{\rm a},j}$, 
is obtained by truncating the expansion of $\omega$ at order $\delta^j$.
To evaluate the accuracy of these approximations, 
we directly compute the resonant frequencies $\omega$ of nearby resonant states with
$\beta=\beta_*+\delta/L$
and define the error
${\rm err}_j=(\omega-\omega_{{\rm a},j})L/2\pi$.
For the at-$\Gamma$ BIC, 
reflection symmetry in $y$ gives the perturbation fields definite parities 
and implies that $\omega_{2\ell+1}=0$ for $\ell\geq 0$. 
Therefore, only approximations of even orders are considered for this BIC.

\begin{figure}[htbp]
	\centering
	\includegraphics[scale=0.33]{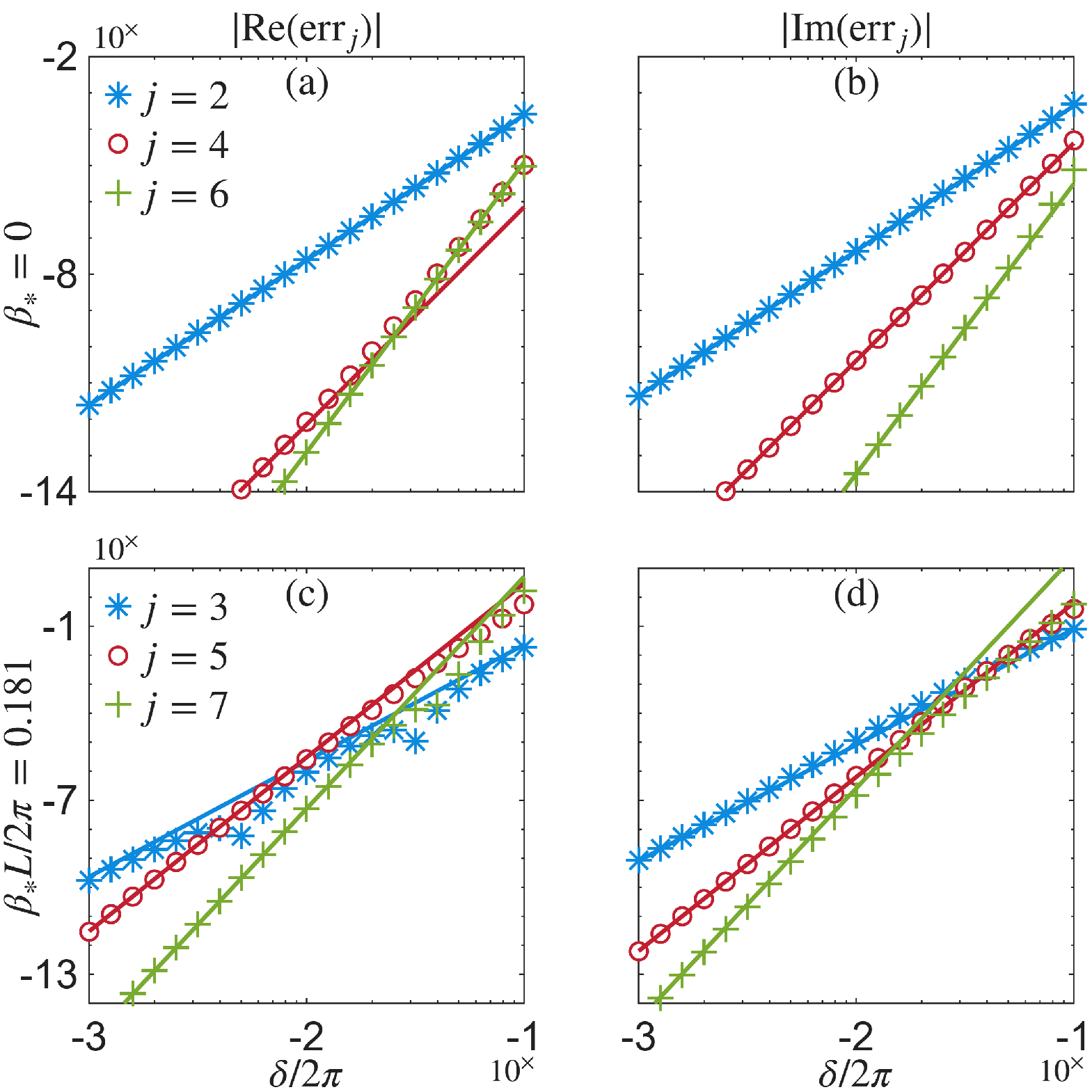}
	\caption{Absolute errors in the real and imaginary parts of $\omega$ for resonant states near the at-$\Gamma$ BIC in (a) and (b) and the off-$\Gamma$ BIC in (c) and (d). The markers represent the directly computed errors, while the lines show the asymptotic errors predicted by the first omitted nonzero terms.}
	\label{ex1}
\end{figure}

Figures~\ref{ex1}(a)--(d) show 
$|\opename{Re}({\rm err}_j)|$ and 
$|\opename{Im}({\rm err}_j)|$ 
for resonant states near the two BICs. 
The markers represent the errors obtained using the directly computed 
resonant frequencies, whereas the lines represent the 
asymptotic errors determined by the first omitted nonzero terms 
in the perturbation expansions. 
For the at-$\Gamma$ BIC, the first omitted nonzero term is 
$\omega_{j+2}\delta^{j+2}$, 
whereas for the off-$\Gamma$ BIC, it is 
$\omega_{j+1}\delta^{j+1}$.
As can be seen from the figures, 
the computed errors agree increasingly well with the corresponding asymptotic errors 
as $\delta$ approaches zero, 
confirming the correctness of the proposed algorithm.

\subsection{A super-BIC with ${\mu}=3$ in a PhC slab with a square lattice} 

We next consider a PhC slab consisting of a uniform dielectric slab 
with refractive index 3.5, 
perforated by a square lattice of circular air holes of radius $0.3L$, 
where $L$ is the lattice period.
We first search for an at-$\Gamma$ super-BIC with $\mu=3$, for which
$Q\sim Q_6(\theta)/\delta^6$ in all directions $\theta$.
This requires the first- and second-order radiation patterns,
${\bm e}_1^\pm(\theta)$ and ${\bm e}_2^\pm(\theta)$,
to vanish for every direction.
For a symmetry-protected BIC, however, if
${\bm e}_1^\pm(\theta)=0$ for all $\theta$,
then ${\bm e}_2^\pm(\theta)=0$ follows automatically from the symmetry constraints~\cite{Zhang25PRL}.
In addition, since the structure is mirror symmetric with respect to $z$,
the upper- and lower-half-space radiation amplitudes
${\bm e}_1^+$ and ${\bm e}_1^-$ are related.
We therefore consider only the upper half-space
and omit the superscript ``$+$'' hereafter.

\begin{figure}[htbp]
	\centering
	\includegraphics[scale=0.33]{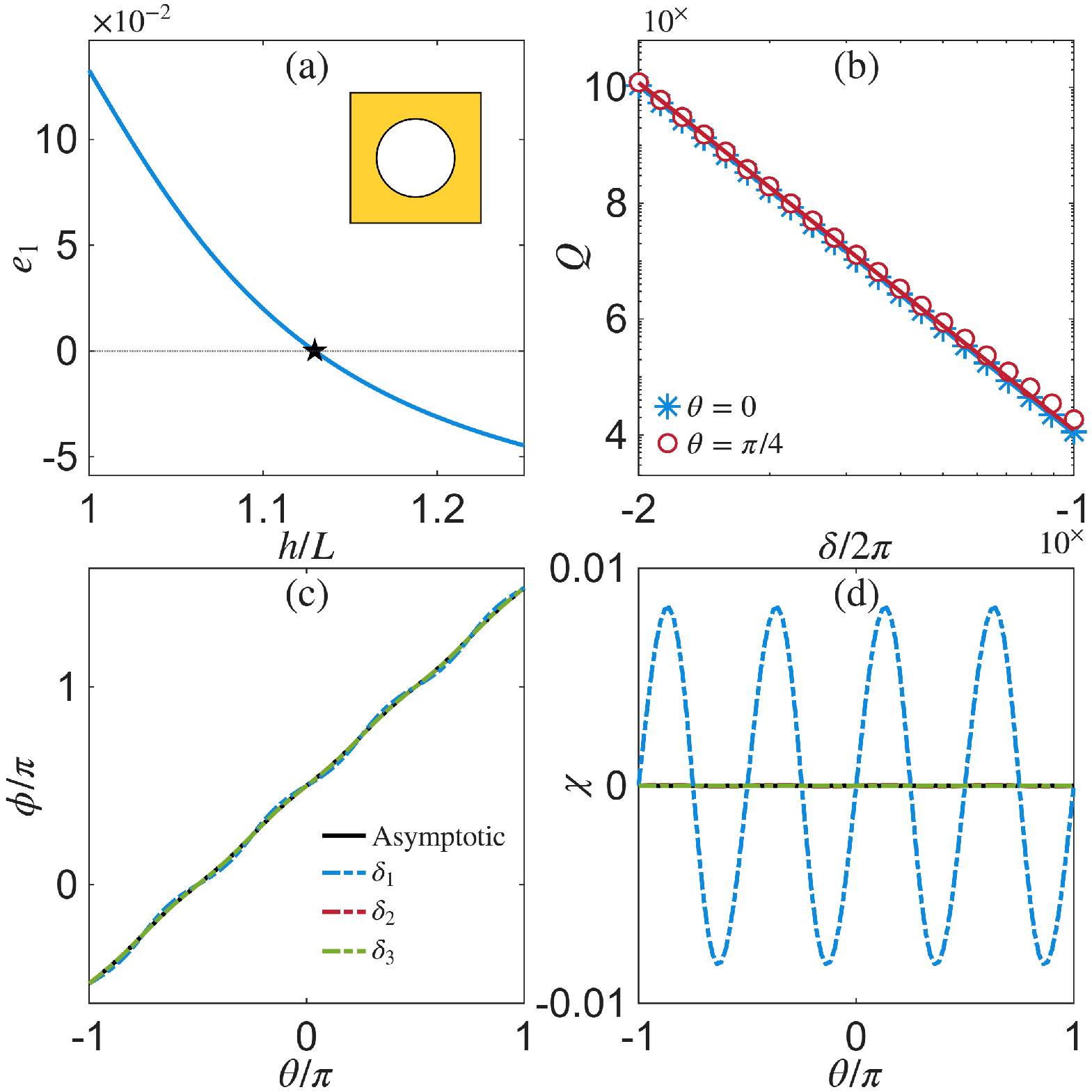}
	\caption{
		Search for and validation of a super-BIC in the PhC slab.
		(a) First-order radiation coefficient $e_1$ as a function of
		the slab thickness $h$ along the family of BICs.
		(b) Directly computed $Q$ factors (open circles) of nearby
		resonant states and the asymptotic approximations
		$Q_6({\theta})/\delta^6$ (lines) for $\theta=0$ and $\pi/4$.
		The two asymptotic lines nearly coincide.
(c), (d) Directly computed polarization angle $\phi$ and DoCP
$\chi$, respectively, for $\delta_i=2\pi\times10^{-i}$,
$i=1,2,3$, compared with the asymptotic approximations $\phi_3$ and $\chi_3$
(labeled ``Asymptotic'') derived from ${\bm e}_3(\theta)$.
	}
	\label{ex2}
\end{figure}

To locate such a state, we compute a family of symmetry-protected
at-$\Gamma$ BICs belonging to the {\em A}$_2$ representation 
by varying the slab thickness $h$ and evaluate
${\bm e}_1({\theta})$ for each BIC.
The rotational symmetry of the structure, together with
Eq.~\eqref{eq:ehmuthetahomo}, implies that if
${\bm e}_1(\theta)=0$ along one direction, then it vanishes along all
directions. We therefore evaluate ${\bm e}_1$ at $\theta=0$,
corresponding to ${\bm \theta}=\hat{x}$. Owing to the reflection
symmetry of the structure, only the $y$ component of ${\bm e}_1$ can
be nonzero. We denote this component by $e_1$ and choose the phase of
${\bm E}_*$ such that $e_1$ is real.
Figure~\ref{ex2}(a) shows $e_1$ as a function of $h$.
The coefficient $e_1$ vanishes at $h=1.13L$, indicating the presence
of a super-BIC with
$\omega_*L/2\pi=0.333$.
For this super-BIC, we solve the perturbation equations up to third order
and obtain the radiation amplitude ${\bm e}_3({\theta})$.
According to Eq.~\eqref{eq:ehmuthetahomo}, it is sufficient to compute
${\bm e}_3$ only at $\theta=0$ and $\pi/4$, since the structure has
$C_4$ rotational symmetry and ${\bm e}_3(\theta)$ is a homogeneous
cubic polynomial in the components of ${\bm \theta}$.

We determine $Q_6({\theta})$ from
${\bm e}_3({\theta})$ and compare the directly computed $Q$ factors
of nearby resonant states with the asymptotic approximation
$Q_6({\theta})/\delta^6$ for $\theta=0$ and $\pi/4$.
The results are shown in Fig.~\ref{ex2}(b).
As $\delta$ approaches zero, the directly computed $Q$ factors agree
closely with the asymptotic predictions, confirming both the identified
super-BIC and the calculated values of $Q_6({\theta})$.

We next verify that ${\bm e}_3({\theta})$ also governs the asymptotic behavior
of the polarization of the nearby resonant states.
Let ${\bm e}_{\bot}=(e_x,e_y)$ denote the projected outgoing radiation
amplitude of a nearby resonant state.
The corresponding Stokes parameters are defined by
\begin{equation}
	\begin{aligned}
		{\cal S}_0
		&=
		|e_x|^2+|e_y|^2,
		&
		{\cal S}_1
		&=
		|e_x|^2-|e_y|^2,
		\\
		{\cal S}_2
		&=
		2\opename{Re}(e_x\overline{e_y}),
		&
		{\cal S}_3
		&=
		-2\opename{Im}(e_x\overline{e_y}).
	\end{aligned}
\end{equation}
We further define the polarization angle $\phi$ and the degree of
circular polarization (DoCP) $\chi$ by
\begin{equation}
	\phi
	=
	\frac{1}{2}
	\arg\left({\cal S}_1+i{\cal S}_2\right),
	\quad
	\chi
	=
	\frac{{\cal S}_3}{{\cal S}_0}.
\end{equation}
The third-order asymptotic approximations $\phi_3$ and $\chi_3$ are obtained by
replacing ${\bm e}_{\bot}$ with
${\bm e}_3=(e_{3x},e_{3y})$ in the definitions above.
Panels (c) and (d) of Fig.~\ref{ex2} compare the third-order asymptotic
approximations with the directly computed polarization quantities at
$\delta_i=2\pi\times10^{-i}$, $i=1,2,3$.
The agreement between $(\phi_3,\chi_3)$ and $(\phi,\chi)$ improves as $\delta$ decreases, 
demonstrating that the third-order perturbation 
solution accurately captures the polarization properties of nearby resonant states. 
Moreover, $\chi_3=0$ for all ${\theta}$, 
showing that the leading radiation near the super-BIC is linearly polarized in every direction, 
consistent with the $C_{4}$ symmetry of the structure. 

\subsection{A super-BIC with $\mu=2$ in a PhC slab with a triangular lattice}

\begin{figure}[htbp]
	\centering
	\includegraphics[scale=0.33]{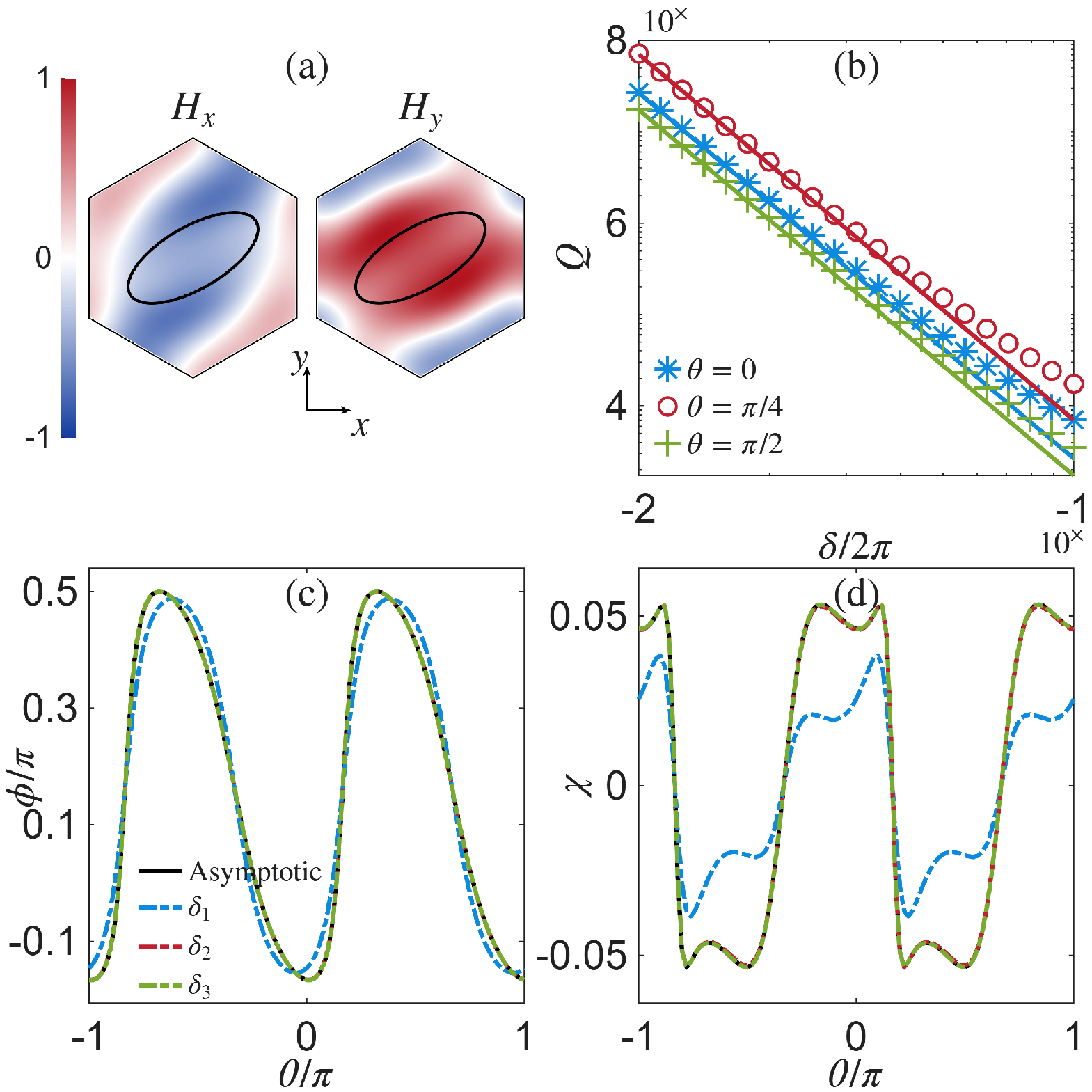}
	\caption{
		Validation of the second-order perturbation solution for the
		at-$\Gamma$ super-BIC with ${\mu}=2$.
		(a) Transverse magnetic-field components of the BIC at $z=0$ within a Wigner--Seitz cell.
		(b) Directly computed $Q$ factors (markers) and the
		asymptotic approximations
		$Q_4({\theta})/\delta^4$ (lines) for
		$\theta=0$, $\pi/4$, and $\pi/2$.
(c), (d) Directly computed polarization angle $\phi$ and DoCP
$\chi$, respectively, for $\delta_i=2\pi\times10^{-i}$,
$i=1,2,3$, compared with the asymptotic approximations $\phi_2$ and $\chi_2$
(labeled ``Asymptotic'') derived from ${\bm e}_2(\theta)$.
	}
	\label{ex3}
\end{figure}

Finally, we consider a uniform dielectric slab perforated by a triangular
lattice of elliptic air holes with lattice constant $L$.
The refractive index of the slab is $3.5$.
The semi-major and semi-minor axes of each hole are $0.35L$ and $0.15L$,
respectively, and the semi-major axis forms an angle of $\pi/6$ with
the $x$ axis.
We search for an at-$\Gamma$ super-BIC with $\mu=2$.
We have shown that non-symmetry-protected at-$\Gamma$ BICs automatically satisfy
${\bm e}_1({\theta})=0$ for all directions ${\theta}$~\cite{Zhang25PRL}.
We therefore vary the structural parameter $h$ to locate such a BIC.
At $h=0.993L$, we find such an at-$\Gamma$ super-BIC with
$\omega_*L/(2\pi)=0.495$.
The transverse magnetic-field components at $z=0$
within a Wigner--Seitz cell are shown in Fig.~\ref{ex3}(a),
confirming the symmetry properties of the mode.
Depending on the structural perturbation preserving the inversion symmetry,
this BIC may either be destroyed or split into a pair of generic
off-$\Gamma$ BICs~\cite{Zhang2024OL}.

We compute the perturbation solution up to second order.
The leading radiation amplitude ${\bm e}_2({\theta})$ is a
homogeneous quadratic polynomial in the components of ${\bm \theta}$.
Using the symmetry constraints, its angular dependence can be recovered
from the values calculated at $\theta=0$, $\pi/4$ and $\pi/2$.
We then obtain $Q_4({\theta})$ from
${\bm e}_2({\theta})$.
Figure~\ref{ex3}(b) compares the directly computed $Q$ factors with
$Q_4({\theta})/\delta^4$ for $\theta=0$, $\pi/4$, and $\pi/2$.
Their agreement as $\delta$ decreases verifies the predicted
fourth-order scaling.

The corresponding polarization approximations $\phi_2$ and $\chi_2$
are obtained by replacing ${\bm e}_{\bot}$ with
${\bm e}_2=(e_{2x},e_{2y})$ in the definitions above.
Panels (c) and (d) of Fig.~\ref{ex3} compare them with the directly
computed polarization quantities for
$\delta_i=2\pi\times10^{-i}$, $i=1,2,3$.
The results converge as $\delta$ decreases, confirming the accuracy
of the second-order perturbation solution.
Unlike in the previous example, however, 
the leading radiation is not constrained to be linearly polarized. 
In fact, structures of this type can support 
resonant states with nearly circular polarization near the BIC~\cite{Zhang2025ACP}.

\section{Conclusion}

Perturbation equations provide not only a rigorous theoretical
description of resonant states near a BIC, but also substantial
numerical advantages. By solving only a few perturbation equations, one
can capture the asymptotic behavior of nearby resonances and
obtain accurate approximations to their frequencies and fields, whereas
direct calculations generally require resonance searches at many nearby
Bloch wavevectors, often along multiple directions in momentum space.
Moreover, when the parameter $\delta$ is sufficiently small, the
imaginary part of the resonant frequency can become extremely small,
making high-accuracy computation of the resonant state increasingly
difficult. By contrast, the perturbation equations are standard
inhomogeneous equations that do not involve the small parameter
$\delta$, and therefore do not suffer from this numerical difficulty.

For layered periodic structures, in which BICs are widely encountered,
we have developed an efficient FMM-based framework for solving the
perturbation equations directly and systematically.
The formulation converts the perturbation equations in each layer into
source-driven Fourier-space ODEs and represents their solutions through
modal expansions with coefficient matrices.
A key feature of the method is that the recursion among the perturbation
equations is transformed into a recursion among polynomial coefficient
matrices.
This structure enables the solutions to be constructed efficiently
and stably to arbitrary order.
The scattering-matrix formulation then couples the solutions across
multiple layers.
The frequency corrections can be obtained either from an analytical
solvability condition or from its discrete Fredholm counterpart.

Numerical examples for both at-$\Gamma$ and off-$\Gamma$ BICs confirm
that the method accurately reproduces the asymptotic behavior of nearby
resonant states. In particular, it resolves the higher-order asymptotic behavior for super-BICs.
The framework can therefore serve not only as a practical tool for
perturbation analysis, but also for identifying and characterizing
special BICs, including chiral BICs and super-BICs.
More broadly, the same source-driven FMM strategy may be useful for
other inhomogeneous electromagnetic problems, including nonlinear
optical calculations.

\appendix

%-------------------------------------------------------------
\section{Derivation of Eq.~\eqref{eq:Q2mu}}
\label{appe:derEqQasym}
%-------------------------------------------------------------

In this appendix, we derive Eq.~\eqref{eq:Q2mu}.
Let $\Omega_h$ denote a unit cell truncated by the planes
$z=\pm h$, where $h$ is sufficiently large that both planes lie
in the homogeneous media, and let $S_h^+$ and $S_h^-$ denote its
upper and lower boundaries, respectively.
For a resonant state, define
\begin{equation}
	{\cal I}_h
	=
	\frac{1}{4}
	\int_{\Omega_h}
	\left[
	{\bm E}^{\dagger}{\bm\varepsilon}{\bm E}
	+
	{\bm H}^{\dagger}{\bm H}
	\right]
	\,{\rm d}{\bm r},
	\label{appeq:Uh}
\end{equation}
and
\begin{equation}
	{\cal P}_h
	=
	\frac{1}{2}
	\opename{Re}
	\int_{S_h^+\cup S_h^-}
	{\bm n}\cdot
	\left(
	{\bm E}\times\overline{\bm H}
	\right)
	\,{\rm d}S,
	\label{appeq:Ph}
\end{equation}
where ${\bm n}$ is the outward unit normal.
Using
\begin{equation}
	\nabla\cdot
	\left(
	{\bm E}\times\overline{\bm H}
	\right)
	=
	\overline{\bm H}\cdot
	(\nabla\times{\bm E})
	-
	{\bm E}\cdot
	(\nabla\times\overline{\bm H}),
\end{equation}
integrating over $\Omega_h$, and taking the real part gives
\begin{equation}
	-2\opename{Im}(\omega){\cal I}_h
	=
	{\cal P}_h.
	\label{appeq:energyid}
\end{equation}
The contributions from opposite lateral boundaries cancel because
of the quasi-periodicity of the resonant field.

Since a resonant field generally grows exponentially as
$|z|\to\infty$ for fixed $\delta\neq0$, we first keep $h$ fixed
and expand both sides of Eq.~\eqref{appeq:energyid} in powers of
$\delta$. The limit $h\to\infty$ is taken only after the
coefficients of equal powers of $\delta$ have been compared.
For fixed $h$, we write
\begin{equation}
	{\cal I}_h
	=
	\sum_{j=0}^{\infty}
	\delta^j{\cal I}_{j,h},
	\quad
	{\cal P}_h
	=
	\sum_{j=0}^{\infty}
	\delta^j{\cal P}_{j,h}.
	\label{appeq:UPexpansion}
\end{equation}

Using the frequency expansion~\eqref{eq:omegaExpansion},
comparison of the coefficient
of $\delta^j$ in Eq.~\eqref{appeq:energyid} gives
\begin{equation}
	-2
	\sum_{\ell=1}^{j}
	\opename{Im}(\omega_\ell)
	{\cal I}_{j-\ell,h}
	=
	{\cal P}_{j,h}.
	\label{appeq:orderjid}
\end{equation}
For $j<2\mu$, every term contributing to
${\cal P}_{j,h}$ contains at least one perturbation field of order
less than $\mu$. Since these lower-order fields contain only
evanescent far-field components,
we have ${\cal P}_{j,h}\to 0$ as $h\to \infty$.
At order $\delta^{2\mu}$, the only contribution that remains finite
as $h\to\infty$ is the product of the $\mu$-th order electric and
magnetic radiation fields. Therefore,
\begin{equation}
	\lim_{h\to\infty}
	{\cal P}_{2\mu,h}
	=
	L^2
	\left(
	{\cal P}_\mu^+
	+
	{\cal P}_\mu^-
	\right),
	\label{appeq:leadingflux}
\end{equation}
where ${\cal P}_\mu^\pm$ are defined in Eq.~\eqref{eq:Pmu}.
The normalization of the BIC,
together with Eq.~\eqref{eq:fieldInnerProduct}, implies
${\cal I}_{0,h}\to L^3$ as $h\to\infty$.
For $j=1$, taking $h\to\infty$ in
Eq.~\eqref{appeq:orderjid} gives
$\opename{Im}(\omega_1)=0$.
Applying the same argument successively yields
$\opename{Im}(\omega_j)=0$ for $1\leq j<2\mu$.
At order $\delta^{2\mu}$, all terms in
Eq.~\eqref{appeq:orderjid} involving
$\opename{Im}(\omega_j)$ with $j<2\mu$ vanish.
Taking $h\to\infty$ and using
Eq.~\eqref{appeq:leadingflux}, we obtain
\begin{equation}
	\opename{Im}(\omega_{2\mu})
	=
	-\frac{
		{\cal P}_\mu^+
		+
		{\cal P}_\mu^-
	}{2L}.
	\label{appeq:Imomega2mu}
\end{equation}
Equation~\eqref{eq:Q2mu} then follows directly.

%--------------------------------------------------------------------
\section{Explicit Forms of Matrices in Secs.~\Rmnum{3} and \Rmnum{4}}
\label{appe:explicitofmatrices}
%--------------------------------------------------------------------

In this appendix, we give the explicit forms of the matrices introduced
in Secs.~\Rmnum{3} and \Rmnum{4}.
For simplicity, we restrict the expressions to isotropic media,
i.e., ${\bm\varepsilon}=\varepsilon({\bm r}){\bm I}$.
The corresponding expressions for anisotropic media can be derived
following the same procedure.

We choose the tangential-field vector as
\begin{equation}
	{\bm\Sigma}(z)
	=
	\left[
	\begin{array}{r}
		{\bf H}_x(z)\\
		{\bf H}_y(z)\\
		-{\bf E}_y(z)\\
		{\bf E}_x(z)
	\end{array}
	\right],
\end{equation}
where ${\bf H}_x(z)$ denotes the Fourier coefficient vector of the
$x$ component of the magnetic field, and the remaining vectors are
defined analogously. 
The matrix ${\bm \Pi}_p$ is given as
\begin{equation}
	{\bm \Pi}_p
	=
	\frac{1}{\omega}
	\begin{bmatrix}
		{\bf 0}
		&
		\omega^2{\mathcal E}_p-{\bf K}_2
		\\
		\omega^2{\bf I}-{\bf K}_{1,p}
		&
		{\bf 0}
	\end{bmatrix},
\end{equation}
where
\begin{equation}
	\begin{aligned}
		&{\bf K}_{1,p}
		=
		\begin{bmatrix}
			{\bm \beta}
			\widehat{\bm \varepsilon}_p^{-1}
			{\bm \beta}
			&
			-{\bm \beta}
			\widehat{\bm \varepsilon}_p^{-1}
			{\bm \alpha}
			\\
			-{\bm \alpha}
			\widehat{\bm \varepsilon}_p^{-1}
			{\bm \beta}
			&
			{\bm \alpha}
			\widehat{\bm \varepsilon}_p^{-1}
			{\bm \alpha}
		\end{bmatrix},
		\label{eq:K1}
		\\
		&{\bf K}_2
		=
		\begin{bmatrix}
			{\bm \alpha}{\bm \alpha}
			&
			{\bm \alpha}{\bm \beta}
			\\
			{\bm \beta}{\bm \alpha}
			&
			{\bm \beta}{\bm \beta}
		\end{bmatrix},
	\end{aligned}
\end{equation}
and
\begin{equation}
	\begin{aligned}
		&{\bm\alpha}
		=
		\operatorname{diag}
		\left(
		\alpha+G_{1,x},
		\ldots,
		\alpha+G_{N,x}
		\right),\\
		&{\bm\beta}
		=
		\operatorname{diag}
		\left(
		\beta+G_{1,y},
		\ldots,
		\beta+G_{N,y}
		\right).
	\end{aligned}
\end{equation}
The matrix $\widehat{\bm\varepsilon}_p$ denotes the dielectric
convolution matrix. 
The matrix ${\mathcal E}_p$ represents the dielectric operator in
Eq.~\eqref{eq:operatorsAandB}.
Using the Laurent factorization rule adopted in this work, we have
${\mathcal E}_p
=
\operatorname{diag}
\left(
\widehat{\bm\varepsilon}_p,
\widehat{\bm\varepsilon}_p
\right)$.
This factorization rule may lead to reduced accuracy when the
dielectric function is discontinuous in the $xy$ plane. 
More accurate factorization
rules can be found in Refs.~\cite{Li1996,Liu12CPC}. We emphasize,
however, that the choice of factorization rule does not affect the
formulation or numerical stability of our algorithm.

We next give the matrices ${\bf Q}_{{\rm J}, p}$ and ${\bf Q}_{{\rm M},p}$
in Eq.~\eqref{eq:qJMrelation}.
For the electric and magnetic current coefficient vectors
\begin{equation}
	{\bf J}
	=
	\left[
	\begin{array}{c}
		{\bf J}_x\\
		{\bf J}_y\\
		{\bf J}_z
	\end{array}
	\right],
	\quad
	{\bf M}
	=
	\left[
	\begin{array}{c}
		{\bf M}_x\\
		{\bf M}_y\\
		{\bf M}_z
	\end{array}
	\right],
\end{equation}
we have
\begin{equation}
	\begin{aligned}
		&{\bf Q}_{{\rm J},p}
		=
		\begin{bmatrix}
			{\bf 0}
			&
			{\bf I}
			&
			{\bf 0}
			\\
			-{\bf I}
			&
			{\bf 0}
			&
			{\bf 0}
			\\
			{\bf 0}
			&
			{\bf 0}
			&
			-\omega^{-1}
			{\bm\beta}
			\widehat{\bm\varepsilon}_p^{-1}
			\\
			{\bf 0}
			&
			{\bf 0}
			&
			\omega^{-1}
			{\bm\alpha}
			\widehat{\bm\varepsilon}_p^{-1}
		\end{bmatrix},
		\\
		&{\bf Q}_{{\rm M},p}
		=
		\begin{bmatrix}
			{\bf 0}
			&
			{\bf 0}
			&
			\omega^{-1}{\bm\alpha}
			\\
			{\bf 0}
			&
			{\bf 0}
			&
			\omega^{-1}{\bm\beta}
			\\
			-{\bf I}
			&
			{\bf 0}
			&
			{\bf 0}
			\\
			{\bf 0}
			&
			-{\bf I}
			&
			{\bf 0}
		\end{bmatrix}.
	\end{aligned}
\end{equation}

Let ${\bf C}_{\theta}$ denote the Fourier representation of the
operation ${\bm\theta}\times$.
Explicitly,
\begin{equation}
	{\bf C}_{\theta}
	=
	\left[
	\begin{array}{ccc}
		{\bf 0}
		&
		{\bf 0}
		&
		\sin\theta\,{\bf I}
		\\
		{\bf 0}
		&
		{\bf 0}
		&
		-\cos\theta\,{\bf I}
		\\
		-\sin\theta\,{\bf I}
		&
		\cos\theta\,{\bf I}
		&
		{\bf 0}
	\end{array}
	\right].
	\label{appeq:Ctheta}
\end{equation}
The Fourier representations of ${\cal A}_1$ and ${\cal B}$ are
\begin{equation}
	{\bf A}_1
	=
	\frac{1}{L}
	\left[
	\begin{array}{cc}
		{\bf 0}
		&
		-{\bf C}_{\theta}
		\\
		{\bf C}_{\theta}
		&
		{\bf 0}
	\end{array}
	\right],
	\quad
	{\bf B}
	=
	\left[
	\begin{array}{cc}
		{\mathscr E}_p
		&
		{\bf 0}
		\\
		{\bf 0}
		&
		{\bf I}
	\end{array}
	\right],
	\label{appeq:A1B}
\end{equation}
where ${\mathscr E}_p=\operatorname{diag}\left(\hat{\bm \varepsilon}_p,\hat{\bm \varepsilon}_p,\hat{\bm \varepsilon}_p\right)$.

We finally give the matrices used to reconstruct 
the complete electric and magnetic fields from the modal coefficient matrix. 
In Eq.~\eqref{eq:CJtoEH}, we have
\begin{equation}
\begin{aligned}
	&{\bf T}_{{\rm E},p}
	=
	\left[
	\begin{array}{cccc}
		{\bf 0}
		&
		{\bf 0}
		&
		{\bf 0}
		&
		{\bf I}
		\\
		{\bf 0}
		&
		{\bf 0}
		&
		-{\bf I}
		&
		{\bf 0}
		\\
		\omega^{-1}
		\widehat{\bm\varepsilon}_p^{-1}
		{\bm\beta}
		&
		-\omega^{-1}
		\widehat{\bm\varepsilon}_p^{-1}
		{\bm\alpha}
		&
		{\bf 0}
		&
		{\bf 0}
	\end{array}
	\right]
	{\bf W}_p,\\
	&{\bf T}_{{\rm H},p}
	=
	\left[
	\begin{array}{cccc}
		{\bf I}
		&
		{\bf 0}
		&
		{\bf 0}
		&
		{\bf 0}
		\\
		{\bf 0}
		&
		{\bf I}
		&
		{\bf 0}
		&
		{\bf 0}
		\\
		{\bf 0}
		&
		{\bf 0}
		&
		-\omega^{-1}{\bm\alpha}
		&
		-\omega^{-1}{\bm\beta}
	\end{array}
	\right]
	{\bf W}_p,\\
	&{\bf T}_{{\rm EJ},p}
	=
	-\frac{i}{\omega}
	\operatorname{diag}
	\left(
	{\bf 0},
	{\bf 0},
	\widehat{\bm\varepsilon}_p^{-1}
	\right),
	\\
	&{\bf T}_{{\rm HM},p}
	=
	-\frac{i}{\omega}
	\operatorname{diag}
	\left(
	{\bf 0},
	{\bf 0},
	{\bf I}
	\right).
	\end{aligned}
\end{equation}

\bibliography{fmm_bic}

@article{Neumann1929PZ,
	title={{\"U}ber merkw{\"u}rdige diskrete Eigenwerte},
	author={von Neumann, J.   and Wigner, E. },
	journal={Phys. Z.},
	volume={30},
	pages={465--467},
	year={1929},
}

@article{Bonnet1994MMAS,
	title={Guided waves by electromagnetic gratings and non-uniqueness examples for the diffraction problem},
	author={Bonnet-Bendhia, A.-S. and Starling, F.},
	journal={Math. Methods Appl. Sci. },
	year={1994},
	volume={17},
	pages={305--338}
}

@article{Marinica08PRL,
	title={Bound {S}tates in the {C}ontinuum in {P}hotonics},
	author={Marinica, D. C. and Borisov, A. G. and Shabanov, S. V.},
	journal={\prl},
	volume={100},
	number={18},
	pages={183902},
	year={2008},
	publisher={APS}
}

@article{Bulgakov08PRB,
	title={Bound states in the continuum in photonic waveguides inspired by defects},
	author={Bulgakov, E. N. and Sadreev, A. F.},
	journal={\prb},
	volume={78},
	number={7},
	pages={075105},
	year={2008},
	publisher={APS}
}

@article{Plotnik11PRL,
	title={Experimental {O}bservation of {O}ptical {B}ound {S}tates in the {C}ontinuum},
	author={Plotnik, Y. and Peleg, O. and Dreisow, F. and Heinrich, M. and Nolte, S. and Szameit, A. and Segev, M.},
	journal={\prl},
	volume={107},
	number={18},
	pages={183901},
	year={2011},
	publisher={APS}
}

@article{Hsu13Nature,
	title={Observation of trapped light within the radiation continuum},
	author={Hsu, Chia Wei and Zhen, Bo and Lee, Jeongwon and Chua, Song-Liang and Johnson, Steven G and Joannopoulos, John D and Solja{\v{c}}i{\'c}, Marin},
	journal={Nature},
	volume={499},
	number={7457},
	pages={188--191},
	year={2013},
	publisher={Nature Publishing Group UK London}
}

@article{Hsu16NRM,
	title={Bound states in the continuum},
	author={Hsu, C. W. and Zhen, B. and Stone, A. D. and Joannopoulos, J. D. and Solja{\v{c}}i{\'c}, M.},
	journal={Nat. Rev. Mater.},
	volume={1},
	pages={16048},
	year={2016},
}

@article{Kivshar2023PU,
	title={Bound states in the continuum in photonic structures},
	author={Koshelev, Kirill Leonidovich and Sadrieva, Zarina Fail'evna and Shcherbakov, Alexey Aleksandrovich and Kivshar, Yurii Semenovich and Bogdanov, Andrey A},
	journal={Phys.-Uspekhi},
	volume={93},
	pages={528--553},
	year={2023}
}

@article{Zhen14PRL,
	title={Topological {N}ature of {O}ptical {B}ound {S}tates in the {C}ontinuum},
	author={Zhen, B. and Hsu, C. W. and Lu, L. and Stone, A. D. and Solja{\v{c}}i{\'c}, M.},
	journal={\prl},
	volume={113},
	number={25},
	pages={257401},
	year={2014},
	publisher={APS}
}

@article{Webster07PTL,
	title={Width dependence of inherent {TM}-mode lateral leakage loss in silicon-on-insulator ridge waveguides},
	author={Webster, M. A. and Pafchek, R. M. and Mitchell, A. and Koch, T. L.},
	journal={IEEE Photonics Technol. Lett.},
	volume={19},
	number={6},
	pages={429--431},
	year={2007},
	publisher={IEEE}
}

@article{Zou15LPR,
	title={Guiding light through optical bound states in the continuum for ultrahigh-{$Q$} microresonators},
	author={Zou, C.-L. and Cui, J.-M. and Sun, F.-W. and Xiong, X. and Zou, X.-B. and Han, Z.-F. and Guo, G.-C.},
	journal={\lpr},
	volume={9},
	number={1},
	pages={114--119},
	year={2015},
	publisher={Wiley Online Library}
}

@article{Zhang24OE,
	title={Non-generic bound states in the continuum in waveguides with lateral leakage channels},
	author={Zhang, Nan and Lu, Ya Yan},
	journal={\opex},
	volume={32},
	number={3},
	pages={3764--3778},
	year={2024},
	publisher={Optica Publishing Group}
}

@article{Gomis17NP,
	title={Anisotropy-induced photonic bound states in the continuum},
	author={Gomis-Bresco, J.  and Artigas, D. and Torner, L.},
	journal={\np},
	volume={11},
	number={4},
	pages={232--236},
	year={2017},
	publisher={Nature Publishing Group}
}

@article{Lee12PRL,
	title={Observation and {D}ifferentiation of {U}nique {H}igh-{$Q$} {O}ptical {R}esonances {N}ear {Z}ero {W}ave {V}ector in {M}acroscopic {P}hotonic {C}rystal {S}labs},
	author={Lee, J. and Zhen, B. and Chua, S.-L. and Qiu, W. and Joannopoulos, J. D. and Solja{\v{c}}i{\'c}, M. and Shapira, O.},
	journal={\prl},
	volume={109},
	number={6},
	pages={067401},
	year={2012},
	publisher={APS}
}

@article{Bulgakov17PRATopo,
	title={Bound states in the continuum and polarization singularities in periodic arrays of dielectric rods},
	author={Bulgakov, E. N. and Maksimov, D. N.},
	journal={\pra},
	volume={96},
	number={6},
	pages={063833},
	year={2017},
	publisher={APS}
}

@article{Kodigala17Nature,
	title={Lasing action from photonic bound states in continuum},
	author={Kodigala, A. and Lepetit, T. and Gu, Q. and Bahari, B. and Fainman, Y. and Kant{\'e}, B.},
	journal={Nature},
	volume={541},
	number={7636},
	pages={196--199},
	year={2017},
	publisher={Nature Publishing Group}
}

@article{Rong23NM,
	title={Spin-valley {R}ashba monolayer laser},
	author={Rong, Kexiu and Duan, Xiaoyang and Wang, Bo and Reichenberg, Dror and Cohen, Assael and Liu, Chieh-li and Mohapatra, Pranab K and Patsha, Avinash and Gorovoy, Vladi and Mukherjee, Subhrajit and Kleiner, Vladimir  and Ismach, Ariel and Koren, Elad  and Hasman,  Erez },
	journal={Nat. Mater.},
	volume={22},
	number={9},
	pages={1085--1093},
	year={2023},
	publisher={Nature Publishing Group UK London}
}

@article{Maria22Science,
	title={Resonant metasurfaces for generating complex quantum states},
	author={Santiago-Cruz, Tom{\'a}s and Gennaro, Sylvain D and Mitrofanov, Oleg and Addamane, Sadhvikas and Reno, John and Brener, Igal and Chekhova, Maria V},
	journal={Science},
	volume={377},
	number={6609},
	pages={991--995},
	year={2022},
	publisher={American Association for the Advancement of Science}
}

@article{Rivera23PNAS,
	title={Creating large {F}ock states and massively squeezed states in optics using systems with nonlinear bound states in the continuum},
	author={Rivera, Nicholas and Sloan, Jamison and Salamin, Yannick and Joannopoulos, John D and Solja{\v{c}}i{\'c}, Marin},
	journal={Proc. Natl. Acad. Sci. U.S.A.},
	volume={120},
	number={9},
	pages={e2219208120},
	year={2023},
	publisher={National Academy of Sciences}
}

@article{JZi20NP,
	title={Generating optical vortex beams by momentum-space polarization vortices centred at bound states in the continuum},
	author={Wang, Bo and Liu, Wenzhe and Zhao, Maoxiong and Wang, Jiajun and Zhang, Yiwen and Chen, Ang and Guan, Fang and Liu, Xiaohan and Shi, Lei and Zi, Jian},
	journal={\np},
	volume={14},
	number={10},
	pages={623--628},
	year={2020},
	publisher={Nature Publishing Group UK London}
}

@article{Yuri20PRL,
	title={Metasurfaces with {M}aximum {C}hirality {E}mpowered by {B}ound {S}tates in the {C}ontinuum},
	author={Gorkunov, Maxim V and Antonov, Alexander A and Kivshar, Yuri S},
	journal={\prl},
	volume={125},
	number={9},
	pages={093903},
	year={2020},
	publisher={APS}
}

@article{Alu21PRL,
	title={Chiral {Q}uasi-{B}ound {S}tates in the {C}ontinuum},
	author={Overvig, Adam and Yu, Nanfang and Al{\`u}, Andrea},
	journal={\prl},
	volume={126},
	number={7},
	pages={073001},
	year={2021},
	publisher={APS}
}

@article{Pura24,
	title={Superchiral light emerging from bound states in the continuum in metasurfaces of {S}i nanorod dimers},
	author={Pura, Jose Luis and Castillo L{\'o}pez de Larrinzar, Beatriz and Liang, Minpeng and Garc{\'\i}a-Mart{\'\i}n, Antonio and G{\'o}mez Rivas, Jaime and S{\'a}nchez-Gil, Jos{\'e} A},
	journal={ACS Photonics},
	volume={11},
	number={10},
	pages={4090--4100},
	year={2024},
	publisher={ACS Publications}
}

@article{Yuri22Science,
	title={Chiral emission from resonant metasurfaces},
	author={Zhang, Xudong and Liu, Yilin and Han, Jiecai and Kivshar, Yuri and Song, Qinghai},
	journal={Science},
	volume={377},
	number={6611},
	pages={1215--1218},
	year={2022},
	publisher={American Association for the Advancement of Science}
}

@article{Friedrich85PRA,
	title = {Interfering resonances and bound states in the continuum},
	author = {Friedrich, H. and Wintgen, D.},
	journal = {\pra},
	volume = {32},
	issue = {6},
	pages = {3231--3242},
	numpages = {0},
	year = {1985},
	publisher = {American Physical Society},
}

@article{Yu26SIAMAM,
	title={Existence of {F}riedrich-{W}intgen bound states in the continuum: system of {S}chr{\"o}dinger dinger equations},
	author={Yu, Xiuchen and Lu, Ya Yan},
	journal={\siamam},
	volume = {86},
	issue = {4},
	pages = {1654--1671},
	year={2026}
}

@article{Susumu14PRL,
	title={Analytical {P}erspective for {B}ound {S}tates in the {C}ontinuum in {P}hotonic {C}rystal {S}labs},
	author={Yang, Yi and Peng, Chao and Liang, Yong and Li, Zhengbin and Noda, Susumu},
	journal={\prl},
	volume={113},
	number={3},
	pages={037401},
	year={2014},
	publisher={APS}
}

@article{Yuan2017OL,
	title={Bound states in the continuum on periodic structures: perturbation theory and robustness},
	author={Yuan, Lijun and Lu, Ya Yan},
	journal={\ol},
	volume={42},
	number={21},
	pages={4490--4493},
	year={2017},
	publisher={Optica Publishing Group}
}

@article{Kivshar19PRB,
	title={Multipolar origin of bound states in the continuum},
	author={Sadrieva, Zarina and Frizyuk, Kristina and Petrov, Mihail and Kivshar, Yuri and Bogdanov, Andrey},
	journal={\prb},
	volume={100},
	number={11},
	pages={115303},
	year={2019},
	publisher={APS}
}

@article{Yuan2021PRAcond,
  title={Conditional robustness of propagating bound states in the continuum in structures with two-dimensional periodicity},
  author={Yuan, Lijun and Lu, Ya Yan},
  journal={\pra},
  volume={103},
  number={4},
  pages={043507},
  year={2021},
  publisher={APS}
}

@article{Zhang23OE,
	title={Robust and non-robust bound states in the continuum in rotationally symmetric periodic waveguides},
	author={Zhang, Nan and Lu, Ya Yan},
	journal={\opex},
	volume={31},
	number={10},
	pages={15810--15824},
	year={2023},
	publisher={Optica Publishing Group}
}

@article{Bezus24PRA,
	title={Algebraic approach to finding the number of parameters required to obtain a bound state in the continuum},
	author={Bykov, Dmitry A and Bezus, Evgeni A and Mingazov, Albert A and Doskolovich, Leonid L},
	journal={\pra},
	volume={109},
	number={5},
	pages={053525},
	year={2024},
	publisher={APS}
}

@article{Yuan20PRAPert,
	title = {Perturbation theories for symmetry-protected bound states in the continuum on two-dimensional periodic structures}, 
	author = {Yuan, L.  and Lu, Ya Yan}, 
	journal = {\pra},
	pages = {043827},
	volume = {101},
	year = {2020},
}

@article{Zhang25PRL,
	title = {Perturbation {T}heory for {R}esonant {S}tates near a {B}ound {S}tate in the {C}ontinuum},
	author = {Zhang, Nan and Lu, Ya Yan},
	journal = {\prl},
	volume = {134},
	issue = {1},
	pages = {013803},
	numpages = {7},
	year = {2025},
	month = {Jan},
	publisher = {American Physical Society}
}

@article{Yuan17PRA,
	title={Strong resonances on periodic arrays of cylinders and optical bistability with weak incident waves},
	author={Yuan, L. and Lu, Ya Yan},
	journal={\pra},
	volume={95},
	number={2},
	pages={023834},
	year={2017},
	publisher={APS}
}

@article{Yuan2018PRA,
  title={Bound states in the continuum on periodic structures surrounded by strong resonances},
  author={Yuan, Lijun and Lu, Ya Yan},
  journal={\pra},
  volume={97},
  number={4},
  pages={043828},
  year={2018},
  publisher={APS}
}

@article{Zhen19Nature,
	title={Topologically enabled ultrahigh-{$Q$} guided resonances robust to out-of-plane scattering},
	author={Jin, J. and Yin, X. and Ni, L. and Solja{\v{c}}i{\'c}, M. and Zhen, B. and Peng, C.},
	journal={Nature},
	volume={574},
	number={7779},
	pages={501--504},
	year={2019},
	publisher={Nature Publishing Group UK London}
}

@article{Bulgakov23PRB, 
	title = {Super-bound states in the continuum through merging in grating}, 
	author = {E. Bulgakov and G. Shadrina and A. Sadreev and K. Pichugin}, 
	journal = {\prb},
	pages = {125303},
	volume = {108},
	year = {2023},
}

@article{Lee2023LPR,
	title={Merging and Band Transition of Bound States in the Continuum in Leaky-Mode Photonic Lattices},
	author={Lee, Sun-Goo and Kim, Seong-Han and Lee, Wook-Jae},
	journal={\lpr},
	volume={17},
	number={12},
	pages={2300550},
	year={2023},
	publisher={Wiley Online Library}
}

@article{Shubin2023PRB,
	title={Interacting bound states in the continuum in {F}abry-{P}{\'e}rot resonators: {M}erging, crossing, and avoided crossing},
	author={Shubin, NM and Kapaev, VV and Gorbatsevich, AA},
	journal={\prb},
	volume={108},
	number={19},
	pages={195419},
	year={2023},
	publisher={APS}
}

@article{Le2024PRL,
	title={Super {B}ound {S}tates in the {C}ontinuum on a {P}hotonic {F}latband: {C}oncept, {E}xperimental {R}ealization, and {O}ptical {T}rapping {D}emonstration},
	author={Le, Ngoc Duc and Bouteyre, Paul and Kheir-Aldine, Ali and Dubois, Florian and Cueff, S{\'e}bastien and Berguiga, Lotfi and Letartre, Xavier and Viktorovitch, Pierre and Benyattou, Taha and Nguyen, Hai Son},
	journal={\prl},
	volume={132},
	number={17},
	pages={173802},
	year={2024},
	publisher={APS}
}

@article{Zhang2025Super,
	title={Super bound states in the continuum: {A}nalytic framework, parametric dependence, and fast direct computation},
	author={Zhang, Nan and Lu, Ya Yan},
	journal={arXiv:2505.00235},
	year={2025}
}

@article{Yuri21NC,
	title={Ultralow-threshold laser using super-bound states in the continuum},
	author={Hwang, M.-S. and Lee, H.-C. and Kim, K.-H. and Jeong, K.-Y. and Kwon, S.-H. and Koshelev, K. and Kivshar, Y. and Park, H.-G.},
	journal={\nc},
	volume={12},
	number={1},
	pages={4135},
	year={2021},
}

@article{Cui25NP,
	title={Ultracompact multibound-state-assisted flat-band lasers},
	author={Cui, Jieyuan and Han, Song and Zhu, Bofeng and Wang, Chongwu and Chua, Yunda and Wang, Qian and Li, Lianhe and Davies, Alexander Giles and Linfield, Edmund Harold and Wang, Qi Jie},
	journal={\np},
	volume={19},
	number={6},
	pages={643--649},
	year={2025},
	publisher={Nature Publishing Group}
}

@book{monk2003finite,
	title={Finite element methods for Maxwell's equations},
	author={Monk, Peter},
	volume={562},
	year={2003},
	publisher={Clarendon Press Oxford}
}

@article{Zhang26JCP,
	title = {Efficient method for computing resonant modes in periodic photonic structures via transverse impedance operators},
	journal = {\jcp},
	volume = {563},
	pages = {115142},
	year = {2026},
	author = {Nan Zhang and Ya Yan Lu}
}

@article{Moharam1981,
  title={Rigorous coupled-wave analysis of planar-grating diffraction},
  author={Moharam, MG and Gaylord, Thomas K},
  journal={\josa},
  volume={71},
  number={7},
  pages={811--818},
  year={1981},
  publisher={Optical Society of America}
}

@article{Li1996,
  title={Use of Fourier series in the analysis of discontinuous periodic structures},
  author={Li, Lifeng},
  journal={\josaa},
  volume={13},
  number={9},
  pages={1870--1876},
  year={1996},
  publisher={Optical Society of America}
}

@article{Lalanne1996,
  title={Highly improved convergence of the coupled-wave method for {TM} polarization},
  author={Lalanne, Philippe and Morris, G Michael},
  journal={\josaa},
  volume={13},
  number={4},
  pages={779--784},
  year={1996},
  publisher={Optical Society of America}
}

@article{Granet1996,
  title={Efficient implementation of the coupled-wave method for metallic lamellar gratings in {TM} polarization},
  author={Granet, G and Guizal, Brahim},
  journal={\josaa},
  volume={13},
  number={5},
  pages={1019--1023},
  year={1996},
  publisher={Optical Society of America}
}

@article{Li1996josaa2,
	title={Formulation and comparison of two recursive matrix algorithms for modeling layered diffraction gratings},
	author={Li, Lifeng},
	journal={\josaa},
	volume={13},
	number={5},
	pages={1024--1035},
	year={1996},
	publisher={Optical Society of America}
}

@article{Whittaker1999,
  title={Scattering-matrix treatment of patterned multilayer photonic structures},
  author={Whittaker, D. M. and Culshaw, I. S.},
  journal={\prb},
  volume={60},
  number={4},
  pages={2610},
  year={1999},
  publisher={APS}
}

@article{Salakhova21PRB,
	title={Fourier modal method for moir{\'e} lattices},
	author={Salakhova, Natalia S and Fradkin, Ilia M and Dyakov, Sergey A and Gippius, Nikolay A},
	journal={\prb},
	volume={104},
	number={8},
	pages={085424},
	year={2021},
	publisher={APS}
}

@article{Hao26PRB,
	title={Near-field thermal radiation in time-varying systems via {F}ourier modal method},
	author={Hao, Yun-Chao and Zhang, Yong and Yi, Hong-Liang},
	journal={\prb},
	volume={113},
	number={3},
	pages={035413},
	year={2026},
	publisher={APS}
}

@article{Thomas26PRB,
  title={Fourier modal method for gratings with chiral, nonreciprocal and bianisotropic materials},
  author={Smagin, Ilia and Weiss, Thomas and Dyakov, Sergey},
  journal={\prb},
  volume={113},
  number={23},
  pages={235408},
  year={2026},
  publisher={APS}
}

@article{Nakagawa2002,
  title={Analysis of enhanced second-harmonic generation in periodic nanostructures using modified rigorous coupled-wave analysis in the undepleted-pump approximation},
  author={Nakagawa, Wataru and Tyan, Rong-Chung and Fainman, Yeshaiahu},
  journal={\josaa},
  volume={19},
  number={9},
  pages={1919--1928},
  year={2002},
  publisher={Optical Society of America}
}

@article{Bai2007,
  title={Fourier modal method for the analysis of second-harmonic generation in two-dimensionally periodic structures containing anisotropic materials},
  author={Bai, Benfeng and Turunen, Jari},
  journal={\josab},
  volume={24},
  number={5},
  pages={1105--1112},
  year={2007},
  publisher={Optical Society of America}
}

@article{Paul2010,
  title={A numerical approach for analyzing higher harmonic generation in multilayer nanostructures},
  author={Paul, Thomas and Rockstuhl, Carsten and Lederer, Falk},
  journal={\josab},
  volume={27},
  number={5},
  pages={1118--1130},
  year={2010},
  publisher={Optical Society of America}
}

@article{Thomas2018OE,
  title={Modeling of second-harmonic generation in periodic nanostructures by the {F}ourier modal method with matched coordinates},
  author={Defrance, Josselin and Sch{\"a}ferling, Martin and Weiss, Thomas},
  journal={\opex},
  volume={26},
  number={11},
  pages={13746--13758},
  year={2018},
  publisher={Optical Society of America}
}

@article{Liu12CPC,
	title={S4: A free electromagnetic solver for layered periodic structures},
	author={Liu, Victor and Fan, Shanhui},
	journal={Comput. Phys. Commun.},
	volume={183},
	number={10},
	pages={2233--2244},
	year={2012},
	publisher={Elsevier}
}

@article{Zhang2024OL,
	title={Bifurcation of bound states in the continuum in periodic structures},
	author={Zhang, Nan and Lu, Ya Yan},
	journal={\ol},
	volume={49},
	number={6},
	pages={1461--1464},
	year={2024},
	publisher={Optica Publishing Group}
}

@article{Zhang2025ACP,
	title={Asymptotically circularly polarized bound states in the continuum},
	author={Zhang, Nan and Lu, Ya Yan},
	journal={arXiv:2511.12900},
	year={2025}
}
\end{document}